\documentclass[preprint,12pt]{article}
\usepackage{subcaption}
\usepackage{svg}
\pdfoutput=1
\usepackage{amssymb}
\usepackage{arxiv}
\usepackage[T1]{fontenc}
\usepackage{hyperref}
\usepackage{url}
\usepackage{booktabs}
\usepackage{nicefrac}
\usepackage{microtype}
\usepackage{doi}
\usepackage{amsfonts}
\usepackage{graphicx}
\usepackage{epstopdf}
\usepackage{algorithmic}
\usepackage{mathtools}
\usepackage{bm}
\usepackage{tikz}
\usetikzlibrary{arrows,positioning}
\usepackage{pgfplots}
\usepackage{amsopn}
\usepackage{amsmath}
\usepackage{empheq}

\usepackage[scientific-notation=true]{siunitx}

\title{A meshfree solver for coupled bulk-surface problems with self-organizing surface geometry}

\date{}

\author{ \href{https://orcid.org/0000-0003-2915-8920}{\includegraphics[scale=0.06]{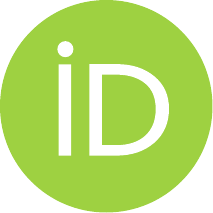}\hspace{1mm}Lennart J.~Schulze\textsuperscript{\textdagger}}\\
	\texttt{lschulze@mpi-cbg.de}
	\And	
    \href{https://orcid.org/0009-0002-3467-2668}{\includegraphics[scale=0.06]{orcid.pdf}\hspace{1mm}Alejandra Foggia\textsuperscript{\textdagger,\textdaggerdbl}}\\
	\texttt{foggia@mpi-cbg.de}
	\And
	\href{https://orcid.org/0000-0003-4414-4340}{\includegraphics[scale=0.06]{orcid.pdf}\hspace{1mm}Ivo F.~Sbalzarini\textsuperscript{*}}\\
	\texttt{ivo.sbalzarini@uzh.ch}} 

\renewcommand{\headeright}{}
\renewcommand{\undertitle}{}
\renewcommand{\shorttitle}{Meshfree Solver for Coupled Bulk-Surface Problems}

\hypersetup{
pdftitle={PDEs on deformable surfaces},
pdfauthor={L.J.~Schulze, A.~Foggia and I.F.~Sbalzarini},
}

\begin{document}
\maketitle
\vspace{-1cm}
\begin{center}
	Dresden University of Technology, Faculty of Computer Science, Dresden, Germany,\\Max Planck Institute of Molecular Cell Biology and Genetics, Dresden, Germany,\\Center for Systems Biology Dresden, Dresden, Germany
\end{center}
\vspace{2cm}
\begingroup
\renewcommand{\thefootnote}{\fnsymbol{footnote}}
\footnotetext[2]{Equal contribution.}
\footnotetext[1]{Corresponding author; now at: Department of Mathematical Modeling and Machine Learning, University of Zurich, Zurich, Switzerland.}
\footnotetext[3]{Now at: Laboratoire Reproduction et D\'eveloppement des Plantes, ENS-Lyon, INRIA, Lyon, France}
\endgroup
\setcounter{footnote}{0}

\begin{abstract}
In many systems, the interaction between a deformable surface or interface and the surrounding bulk fluid is coupled with intrinsic spatiotemporal dynamics within the moving surface. Examples include tumor growth, biological tissue morphogenesis, cardiac mechanics, multi-phase surfactant chemistry, additive manufacturing, clothing wear-and-tear, and reactive combustion flows. Solving such problems requires both geometric computing algorithms to track and resolve the surface and numerical methods to solve the coupled governing equations in the surface and the surrounding bulk phase. We here present a fully meshfree numerical solver for such coupled bulk-surface problems with deformable interfaces. The presented solver tracks the surface implicitly, solving for the dynamic surface geometry based on stress balance coupled to surrounding fluid phases. We show convergence for a mass-conserving case on a growing sphere and solve bulk-surface problems with incompressible Navier--Stokes fluids coupled to in-surface nonlinear reaction-diffusion dynamics. Finally, we show a  model of biological morphogenesis, solving simultaneously for the dynamic surface shape and the fields on the curved surface with two-way coupling.
\end{abstract}

\keywords{bulk-surface coupling \and surface mechanics \and fluid surfaces \and PDEs on surfaces \and particle methods \and geometric computing \and meshfree methods}

\section{Introduction}
In many scientific and engineering applications, surfaces do not only constitute interfaces between two fluid phases, but they represent a curved domain of time-varying shape with own intrinsic dynamics. We here consider the case where the dynamic shape of a surface or interface is not imposed as a boundary condition but emerges from the intrinsic stress balance of coupled in-surface and bulk continuum mechanics. The surface shape, as well as the unknown fields on the surface itself, then co-evolve according to the physics of the surface coupled with the surrounding bulk fluid phases, giving raise to emergent surface dynamics. 

Emergent surface dynamics are key, for example, in the additive manufacturing process of selective laser melting. There, the dynamics within the surface of the melt pool of liquefied metal defines the quality of the resulting manufactured metal parts: Gradients in temperature and surfactants within the surface of the melt pool create surface tension gradients, which in turn drive flows within the surface and change the shape of the surface itself. These in-plane Marangoni flows can lead to undesired outcomes in the manufacturing process \cite{li2022numerical,siao2021examination}. Another example is found in developmental biology, where intrinsic surface dynamics of epithelial tissues plays a key role in organ morphogenesis. In-surface processes include cell division, cell-shape changes, and cell motility, which jointly lead to tissue shape changes including buckling and folding \cite{jain2020regionalized,pinheiro2022morphogen,fuhrmann2024active}. As a special case, surface tension and stresses have been shown to drive heart development, where the surface of the heart changes its shape coupled with the hydrodynamics of the blood and the exterior intersticial fluid \cite{heart_simulation2023,mechanical_fracture_heart2025}. As a final example, intrinsic surface dynamics is also of interest in computer graphics and computer vision, e.g., when generating realistic motion of clothes or elastic deformations and fracture propagation in thin-shell objects \cite{FEM_NS_evolving_surfaces2023}. In all of these examples, the time evolution of the surface shape is not known {\it a priori} and must be determined along with the density, velocity, and pressure fields within the curved surface and in the surrounding bulk phases by solving a coupled set of partial differential equations (PDEs).

Numerically solving coupled bulk-surface PDEs with time-varying surface deformations hinges on accurate computation of the differential geometry of the surface. Besides the actual surface location, which is important for the coupling with the surrounding bulk fluids, this includes the surface normal field and the surface's curvature tensor or metric tensor field as a function of space and time. These constitute higher-order quantities, which must be approximated with sufficiently high order of accuracy to not dominate the error of the PDE solver. In addition, the spatial resolution of the surface discretization has to adapt to surface curvature changes as well as surface-area changes in order to remain well sampled. When approximating surface differential operators, the metric tensor of the surface has to be accounted for, introducing additional curvature terms in the momentum-balance equations, with their own time dynamics. This may render the coupled bulk-surface PDEs numerically stiffer than their flat-space or fixed-shape counterparts, leading to more restrictive Courant--Friedrichs--Lewy (CFL) conditions on the numerical solver. 

Several numerical methods have been proposed for intrinsic surface dynamics. They  conceptually differ mainly in whether the geometry is represented explicitly or implicitly. Finite-element methods (FEM) solve the governing equations in their weak form \cite{dziuk2013finite} and traditionally represent the surface by an explicit mesh \cite{dziuk2006finite}. To allow for deformations of the surface, the surface mesh moves, generating a Lagrangian frame of reference for the PDE on the surface \cite{dziuk2007finite} and a dynamic domain for the coupled bulk-surface problem \cite{elliott2013finite}. To ensure compatibility between the bulk meshes and the time-evolving surface mesh, methods such as TraceFEM \cite{gross2015trace,olshanskii2018trace} and the closely related CutFEM \cite{burman2015cutfem} have been proposed. When using an Eulerian frame of reference, the surface is usually represented implicitly using phase-field \cite{ratz2006pdes} or level-set methods \cite{bertalmio2001variational}. With such implicit surface representations, adaptive finite differences have been used to solve multi-phase flow problems with variable surface tension in the strong form \cite{teigen2011diffuse}.

Strong-form solvers that avoid the technicalities of mesh generation and mesh maintenance in dynamic domains have been constructed using meshfree collocation methods \cite{overview_particle_methods_Halada2025}. Meshfree methods naturally combine a Lagrangian frame of reference with implicit surface representations, such as level sets for PDEs on deforming surfaces~\cite{bergdorf2010lagrangian} and phase-fields for coupled bulk-surface problems~\cite{tauriello2013coupling}. Meshfree collocation methods have also successfully been used with explicit surface representations \cite{suchde2019meshfree,suchde2019fully}. The surface variant of discretization-corrected particle strength exchange (DC-PSE) \cite{schrader2010discretization} relies on a semi-implicit representation by surface point clouds. This enables high-order approximations of surface differential operators using a concept related to the closest point transform \cite{ruuth2008simple}. Surface fields are invariantly extended along the normal to approximate surface differential operators by projections of bulk DC-PSE operators \cite{singh2023meshfree}. Surface DC-PSE therefore naturally couples intrinsic surface dynamics with fields in the embedding space.

Here, we leverage this bulk-surface coupling and the high-order accuracy of surface DC-PSE to construct a solver for coupled bulk-surface dynamics with two-way interaction via deformable surfaces. We compute the required geometric surface quantities, like normals and curvatures, using the particle closest-point (PCP) method \cite{schulze2024high} with high order of accuracy. To ensure robust approximation of surface differential operators even for large deformations of the surface, we periodically resample the surface using the self-adaptive implicit surface sampling (SAISS) method \cite{schulze2026globally}. The resulting solver is entirely meshfree and combines advantages of both explicit and implicit surface representations: The implicit polynomial level-set representation of PCP enables the computation of accurate geometric quantities throughout large deformations and changes in surface topology~\cite{schulze2024high}. The explicit representation by the DC-PSE surface particles accurately tracks the surface in a Lagrangian frame of reference and enables accurate in-surface computations. There is no requirement for conforming surface and bulk particle distributions, simplifying the implementation of the method and contributing to its versatility. Both the surface and bulk particles are advanced in time using an explicit predictor-corrector scheme, providing enhanced stability for potentially stiff coupled bulk-surface PDEs \cite{jiang2025predictor}. The spatially local particle-particle interaction kernels of DC-PSE in combination with explicit time integration enables implementing the presented solver on shared- and distributed-memory parallel computers.

We provide a parallel software implementation using the open-source C++ library OpenFPM for scalable scientific computing \cite{incardona2019openfpm}. All elements of the solver are implemented as transparent OpenFPM data containers, providing performance portability across hardware platforms \cite{Incardona:2023a}. We show convergence of the algorithm for a mass-conserved case on a growing sphere. We then demonstrate the applicability of the presented solver to coupled bulk-surface problems by simulating nonlinear reaction-diffusion patterns on an oscillating droplet, where the surface deformation is the result of two-phase Navier-Stokes equations in the bulk. Finally, we consider a problem with full two-way coupling between the intrinsic surface dynamics and the shape of the surface by simulating the emergence of different organic shapes from a morphogenetic model.

\section{Physical model and governing equations}

We consider a general continuous reaction-diffusion equation within a dynamic surface. Nonlinear reaction-diffusion dynamics can create spatial information by diffusion-driven instabilities known as Turing patterns \cite{turing1952chemical}.
Reaction-diffusion equations model the dynamics of a set of chemical species $\mathbf{c}=c^{[s]}$, $s=\{1,\dots,n_{c}\}$, diffusing independently in space but reacting with each other over time. Here, the domain of this process is a closed dynamic two-dimensional (2D) surface $\Gamma(t)$ embedded in a three-dimensional (3D) bulk domain $\Omega$. This is governed by the PDE
\begin{equation}
    \frac{\mathrm{D}c^{[s]}}{\mathrm{D}t}=D^{[s]}\Delta_{\Gamma} c^{[s]} + R(\mathbf{c}) - c^{[s]}\nabla_\Gamma\cdot\mathbf{u}\qquad~\text{in~}\Gamma(t)\, ,
    \label{eq:appPDE}
\end{equation}
where $D^{[s]}$ is the diffusion constant of species $s$, $\Delta_\Gamma$ the surface-intrinsic Laplace--Beltrami operator, and $\nabla_{\Gamma} = \mathrm{P} \nabla = (\mathrm{I} - \mathbf{n}\otimes\mathbf{n}) \nabla$ the covariant derivative in $\Gamma$~\cite{Stone1990}. The reaction term $R(\mathbf{c})$ is typically nonlinear in the concentrations $\mathbf{c}$, and the last term on the right-hand side accounts for local concentration changes due to surface deformation with velocity $\mathbf{u} = \mathbf{u}_{\Gamma} + \mathbf{u}_{\perp}$. The material (Lagrangian) derivative with respect to the moving and deforming surface is
\begin{equation*}
    \frac{\mathrm{D}c^{[s]}}{\mathrm{D}t}=\frac{\partial c^{[s]}}{\partial t}+\left(\mathbf{u}_\Gamma+\mathbf{u}_\perp\right)\cdot\nabla_{\Gamma} c^{[s]}
    =\frac{\partial c^{[s]}}{\partial t}+\mathbf{u}_\Gamma\cdot\nabla_\Gamma c^{[s]}\underbrace{+\mathbf{u}_\perp\cdot\nabla_\Gamma c^{[s]}}_{=0}=\frac{\partial c^{[s]}}{\partial t}+\mathbf{u}_\Gamma\cdot\nabla_\Gamma c^{[s]}\, ,
\end{equation*}
where the normal component of the velocity, $\mathbf{u}_{\perp}$, and the in-surface gradient of the concentration field, $\nabla_\Gamma c^{[s]}$, are perpendicular by definition. Hence, only deformations in the tangential plane of the surface contribute to the advective transport of the surface concentration field, whereas the normal component of the deformation velocity influence the surface concentration through the last term on the right-hand side of Eq.~\eqref{eq:appPDE}:
\begin{equation*}
    c^{[s]}\nabla_\Gamma\cdot\mathbf{u} = c^{[s]} \nabla_\Gamma \cdot (\mathbf{u}_\Gamma +\mathbf{u}_\perp) = c^{[s]} \nabla_\Gamma \cdot \mathbf{u}_\Gamma + c^{[s]} \nabla_\Gamma \cdot ((\mathbf{u}\cdot \mathbf{n})\mathbf{n}) = c^{[s]} \nabla_\Gamma \cdot \mathbf{u}_\Gamma + c^{[s]}\underbrace{(\nabla_\Gamma \cdot \mathbf{n})}_{=-\kappa} (\mathbf{u}\cdot\mathbf{n})\, ,
\end{equation*}
with $\kappa$ the local mean curvature. In practice, we directly solve for $c^{[s]}\nabla_\Gamma\cdot\mathbf{u}$, as the orthogonal extension of the velocity into the embedding space takes care of the normal contribution.

The deformation velocity of the surface is obtained in one of two ways: (a) through a constitutive law 
\begin{equation}\label{eq:appConstitutiveVel}
    \mathbf{u}(\mathbf{x})=F(\mathbf{c}(\mathbf{x}))\, ,
\end{equation}
or (b) by solving a coupled set of PDEs in the embedding volume $\Omega$ of the surface. We assume the embedding volume to be filled with incompressible fluid phases of viscosity $\eta$, local density $\rho$, and pressure $P$. The two fluid phases are separated by the surface with surface tension $\mathbf{F}^{(s)}$. The velocity field is obtained by solving the Navier--Stokes equations in the embedding space,
\begin{empheq}[left=\left.,right=\right\}\quad\text{in}~\Omega.]
{align}
    &\frac{\mathrm{D}\rho}{\mathrm{D}t}=-\rho\nabla\cdot\mathbf{u}\label{eq:appmasscontieq}\\
    &\frac{\mathrm{D}\mathbf{u}}{\mathrm{D}t}=-\frac{1}{\rho}\nabla P+\frac{1}{\rho}\eta\Delta\mathbf{u}+\mathbf{F}^{(s)}(\kappa,\mathbf{n})\label{eq:appmomentumeq}
\end{empheq}
Here, $\nabla$ and $\Delta$ are the bulk Euclidean Nabla and Laplace operators, respectively. The surface tension acting within the interface of the two phases is modeled by its resultant volumetric surface tension force $\mathbf{F}^{(s)}$, which is proportional to the curvature $\kappa=\kappa_1+\kappa_2$ with principal curvature values $\kappa_{1,2}$ of the surface and acts orthogonally to the surface. We here use the classic continuum surface force (CSF) model \cite{brackbill1992continuum}, $\mathbf{F}^{(s)}=-\frac{\tau}{\rho}\kappa\mathbf{n}\delta_\epsilon$ with a smoothed surface delta function $\delta_\epsilon$ (a compact one-dimensional kernel, see Eq.~\ref{eq:csfmodel}) and a constant tension coefficient $\tau$. The system of PDEs \eqref{eq:appmasscontieq}-\eqref{eq:appmomentumeq}, is closed by the Cole equation of state \cite{cole1948underwater}
\begin{equation}
    P(\rho)=\frac{c_{ss}^2\rho_0}{\gamma}\left(\left(\frac{\rho}{\rho_0}\right)^{\!\gamma}-1\right) \label{eq:appEoS}
\end{equation}
with reference density $\rho_0$, artificial speed of sound $c_{ss}$, and polytropic index $\gamma$. While, in principle, the two fluid phases can have different reference densities $\rho^{[0]}_0,\rho^{[1]}_0$ and viscosities $\eta^{[0]},\eta^{[1]}$ we here use $\rho^{[0]}_0=\rho^{[1]}_0=\rho_0$ and $\eta^{[0]},\eta^{[1]}=\eta$.

Regardless of the model that governs the deformation of the surface, we track the surface and compute surface normals $\mathbf{n}$ and local curvatures $\kappa$ over time by representing $\Gamma(t)$ as the zero level-set of a scalar function $\phi(\mathbf{x},t):\mathbb{R}^{3}\mapsto\mathbb{R}$, such that
\begin{equation}
    \Gamma (t)=\{\mathbf{x}~:~\phi(\mathbf{x},t)=0\}\, .
\end{equation}
Dynamic deformation of the surface by the velocity field $\mathbf{u}$ is modeled through the kinematic boundary condition as an advection of the level-set field
\begin{equation}
    \frac{\mathrm{D}\phi}{\mathrm{D}t}=\frac{\partial\phi}{\partial t}+\mathbf{u}\cdot\nabla\phi=0.
\end{equation}
Since gradients of scalar fields are perpendicular to iso-contours, unit surface normals are computed as
\begin{equation}
    \mathbf{n}(\mathbf{x})=\frac{\nabla\phi(\mathbf{x})}{\|\nabla\phi(\mathbf{x})\|_2}
\end{equation}
and the mean curvature in its fluid-mechanical definition is
\begin{equation}
    \kappa(\mathbf{x})=\nabla\cdot\mathbf{n}(\mathbf{x})\, .
\end{equation}

The PDEs within the surface and the shape dynamics of the surface are fully coupled if the unknown variables of the surface PDE affect the surface deformation, and vice versa. In the above model, this only holds for case (a). 
There, the concentration field within the surface determines the shape change (Eq.~\eqref{eq:appConstitutiveVel}), which in turn affects the dynamics of the concentration field within the surface (Eq.~\eqref{eq:appPDE}). In case (b), 
the bulk hydrodynamics (Eq.~\eqref{eq:appmasscontieq}-\eqref{eq:appmomentumeq}) determine the space in which the intrinsic surface dynamics take place and thereby affect the surface concentration field. There is, however, no feedback from the surface concentration to the bulk hydrodynamics. This could be introduced, e.g., by making the surface tension depend on local surfactant concentration (Marangoni effect). We here use the models in their present form in order to be able to test both one-way and two-way coupling between in-surface fields and surface deformation dynamics.

\section{Numerical method}

We discretize the curved surface and the embedding space using two distinct sets of particles: the surface particle set $\mathcal{P}_s=\{\mathbf{x}_0,\ldots,\mathbf{x}_{i_s},\ldots,\mathbf{x}_{n_s-1}\}$ with average inter-particle spacing (resolution) $h_s$ and particle positions $\mathbf{x}_{i_s}$ for $i_s\in\{0,\ldots,n_{s}-1\}$, and the bulk particle set $\mathcal{P}_b=\{\mathbf{x}_0,\ldots,\mathbf{x}_{i_b},\ldots,\mathbf{x}_{n_b-1}\}$ with average inter-particle spacing $h_b$ and particle positions $\mathbf{x}_{i_b}$ for $i_b\in\{0,\ldots,n_{b}-1\}$. The surface particles are located exclusively on the surface and are used to track the surface and solve the reaction-diffusion Eq.~\eqref{eq:appPDE}. They store the concentrations of the surface species $c^{[s]}$ and the surface normals required to construct surface DC-PSE differential operators (see Sec.~\ref{sec:surfDCPSE}).

The bulk particles are used to store and track the level-set function, enabling accurate computation of differential-geometric quantities using the PCP method (see Sec.~\ref{sec:pcp}). In the presence of a bulk fluid phase, the bulk particles also store the bulk fluid velocity, density, and pressure fields. Then, the dynamics of the embedding fluid phases is solved using two-phase Smoothed Particle Hydrodynamics (SPH) \cite{monaghan2005smoothed} (see Sec.~\ref{sec:sphmultiphaseflow}).

Both sets of particles can freely move with the material, and we compute all differential operators in the Lagrangian frame of reference. The resulting representation is entirely meshfree with a semi-implicit surface representation: the location of the surface particles explicitly represents the surface for the purpose of solving the surface PDEs, while the local level-set description enables accurate differential geometry and coupling with the bulk. As particles move, we ensure well-sampledness and numerical consistency by periodic regulatization of the particle distribution (see Sec.~\ref{sec:saiss}) and use a predictor-corrector time integration scheme (see Sec.~\ref{sec:timeint}).

\subsection{Multi-phase bulk hydrodynamics using SPH}\label{sec:sphmultiphaseflow}

To solve the multi-phase hydrodynamics governed by the Navier--Stokes equations \eqref{eq:appmasscontieq}-\eqref{eq:appmomentumeq}, we use SPH operators \cite{adami2010new} on the bulk particles. These particles store density $\rho_{i_b}$, velocity $\mathbf{u}_{i_b}$, and pressure $P_{i_b}$. Their locations $\mathbf{x}_{i_b}$ are initialized on a regular Cartesian grid with resolution $h_b$. From there, the particles move with the resulting fluid flow to provide a Lagrangian discretization. The densities $\rho_{i_b}$ are computed using density summation,
\begin{equation}\label{eq:densitysummation}
    \rho_{i_b}=m_{i_b}W_{i_bj_b}\, .
\end{equation}
All bulk particles carry the same mass $m_{i_b}=m~\forall i_b$, which is chosen such that $\sum_{i_b}m=V_{b,\text{total}}\rho_0$, with $V_{b,\text{total}}$ the volume occupied by the fluid. The kernel function $W_{i_bj_b}$, modeling pairwise interactions between bulk particles $i_b$ and $j_b$, is the 3D Wendland C2 kernel \cite{wendland1995piecewise} with smoothing length $\epsilon$. This yields the volumes occupied by each bulk particle as
\begin{equation}
    V_{i_b}=\left( \sum_{j_b}W_{i_bj_b}\right) ^{\!-1}.
\end{equation} 
From the density field, the pressure field is computed Eq.~\eqref{eq:appEoS}.
The velocity updates for all particles are computed using the discretized momentum equation
\begin{equation}\label{eq:appDiscreteMomentum}
    \begin{aligned}
        \frac{\mathrm{D}\mathbf{u}_{i_b}}{\mathrm{D}t}=&-\frac{1}{m}\sum_{j_b}\left(V_{i_b}^2+V_{j_b}^2\right)\frac{\rho_{i_b}P_{j_b}+\rho_{j_b}P_{i_b}}{\rho_{i_b}+\rho_{j_b}}\nabla W_{i_bj_b}\\
        &+\frac{1}{m}\sum_{j_b}\eta\left(V_{i_b}^2+V_{j_b}^2\right)\frac{\mathbf{u}_{i_bj_b}}{r_{i_bj_b}}\frac{\partial W}{\partial r_{i_bj_b}}+\mathbf{F}_{i_b}^{(s)}\, ,
    \end{aligned}
\end{equation}
where $r_{i_bj_b}=\|\mathbf{x}_{i_b}-\mathbf{x}_{j_b}\|_2$ and $\mathbf{u}_{i_bj_b}=\mathbf{u}_{i_b}-\mathbf{u}_{j_b}$.
The volumetric surface tension acting on bulk particle $i_b$ in Eq.~\eqref{eq:appDiscreteMomentum} is computed using a 1D Wendland C2 kernel, $W_{1}$, with smoothing length $\epsilon_{1}$ to regularize the Dirac distribution as
\begin{equation}\label{eq:csfmodel}
    \mathbf{F}^{(s)}_{i_b}=-\frac{\tau}{\rho_{i_b}}\kappa_{i_b}\mathbf{n}_{i_b}W_{1}(\phi_{i_b}).
\end{equation}

\subsection{Surface tracking and quantification using PCP}\label{sec:pcp}

The bulk particles at positions $\mathbf{x}_{i_b}$ also store geometric information in form of the level-set values $\phi_{i_b}$ and their respective closest points on the surface, $\mathbf{cp}_{i_b}$. Further, they cache the surface normals $\mathbf{n}_{i_b}$ and mean and Gaussian curvatures $\kappa_{i_b}, \tilde{\kappa}_{i_b}$ at their closest points on the surface --- $\mathbf{n}_{i_b}=\mathbf{n}_{i_b}(\mathbf{cp}_{i_b})$, $\kappa_{i_b}=\kappa_{i_b}(\mathbf{cp}_{i_b})$, $\tilde{\kappa}_{i_b}=\tilde{\kappa}_{i_b}(\mathbf{cp}_{i_b})$. These quantities are only recomputed if the surface has moved.

All geometric quantities, closest surface points, normals, and curvatures, are computed to high order of accuracy using the PCP method \cite{schulze2024high}.
PCP uses local regression polynomials in patches of radius $r_c$ along the surface. These polynomials are constructed in Lagrange basis \cite{hecht2026multivariate}
yielding a continuous representation of the level-set function that is robust to Lagrangian distortion in the particle locations. Since the representation is piecewise polynomial, derivatives are available analytically and everywhere (also between particle locations) for computing normals and curvatures. For all results presented here, we use polynomials of $L_1$-degree four. The other parameters of the PCP method, including the solver tolerance and the size of the cutoff radius, are mentioned in the results section \ref{sec:results}.

The PCP method only requires bulk particles within a narrow band of width $w$ to both sides of the surface (tubular neighborhood in the embedding space). When solving coupled bulk-surface PDEs, as in case (b), we require bulk particles everywhere in the embedding space. We then use the same bulk particles to also store the level-set information within the narrow band. When only solving a PDE within a dynamic surface, as in case (a), bulk particles are limited to the narrow band, and they only store the level-set function values.

\subsection{Surface differential operator approximation using surface DC-PSE} \label{sec:surfDCPSE}

We solve Eq.~\eqref{eq:appPDE} using surface DC-PSE \cite{singh2023meshfree}. For each time step $t^k$, we compute the Laplace--Beltrami operator of the surface concentration fields $c^{[s]}$ and the surface divergence of the deformation velocity $\mathbf{u}$. Surface DC-PSE is based on a similar principle as the closest-point method for the computation of surface differential operators \cite{ruuth2008simple}. Both approaches exploit the fact that intrinsic surface differential operators can be obtained as volumetric differential operators over fields that are constant along the surface normals.

Therefore, surface DC-PSE extends the surface particles along their respective normals to create an auxiliary set of {\em virtual particles} $i_v$. The virtual particles have locations $\mathbf{x}_{i_v}$ evenly spaced along the positive and negative orthogonal direction with inter-particle distance $h_s$. Their values are copies of the values carried by the originating surface particle, hence $f(\mathbf{x}_{i_v})=f(\mathbf{cp}(\mathbf{x}_{i_v}))$ for any field $f$. The number of layers of virtual particles depends on the order of the differential operator and the desired order of convergence. The resulting narrow band must not self-intersect.

This creates orthogonally extended surface concentration fields $c^{[s]}$ and an orthogonally extended deformation velocity field $\mathbf{u}$. Subsequently, bulk DC-PSE \cite{schrader2010discretization} operators of these fields are computed in neighborhoods of cutoff radius $r_\text{DCPSE}$, containing both virtual and real surface particles. This yields consistent approximations of intrinsic surface differential operators of scalar fields to any desired order of convergence \cite{singh2023meshfree}. 

\subsection{Regularization of the surface particle distribution using SAISS}\label{sec:saiss} 

Surface DC-PSE, as other meshfree and particle methods, requires a certain regularity and density in the distribution of the computational nodes. Surface deformations and Lagrangian movement of the particles can lead to violation of these requirements. Tangential surface stretching and shrinking, for example, creates undesired holes and clusters in the surface particle distribution, which eventually lead to numerically inconsistent or unstable discrete operators.

We prevent inconsistencies and instabilities by periodically resampling the surface particle distribution using the SAISS method \cite{schulze2026globally}. SAISS also transparently adds and removes particles, adjusting the overall size of the particle set to the dynamically evolving surface area and dynamically regularizing the particle distribution. The frequency of the regularization step, $\mathfrak{f}_\text{SAISS}$, depends on the dynamics of the surface: Rapidly deforming surfaces require more frequent regularization.

SAISS adds, removes, and redistributes particles using an optimization loop to match the desired resolution $h_s$ everywhere on the surface. We do not use the optional curvature adaptivity provided by SAISS (i.e., $\tau_\text{SAISS}=0$ in Ref.~\cite{schulze2026globally}), yielding intrinsically equidistant point distributions on the curved surface. As a threshold accuracy for the optimization loop in SAISS, we use $\varepsilon_\text{SAISS}=10^{-4}$. The rest of the SAISS parameters are left at their default values: $r_s=2.0h_s$ for the local support measure, lower and upper threshold values for the insertion and removal of particles of 0.8 and 1.2, respectively, and $r^*=2.0h_s$ for the computation of the energy cost and its gradient. See Ref.~\cite{schulze2026globally} for details on these parameters.

After each particle regularization step, the particle properties are interpolated from the old set of particles $\mathcal{S}^-_s$ to the new, regularized set $\mathcal{S}_s^+$. This in-surface interpolation is done using surface DC-PSE operators for the zeroth derivative applied to the surface fields $c^{[s]}$ \cite{reboux2012self}. The orthogonal extension of the virtual particles in surface DC-PSE accounts for the local curvature of the surface, yielding a surface-intrinsic interpolation. For the results shown in this paper, we always use second-order interpolation with the same cutoff radius as for the other surface DC-PSE operators.

\subsection{Time integration for surface-only PDEs}\label{sec:timeintsurfacePDE}

\begin{figure}
    \centering
    \tikzstyle{block} = [rectangle, draw, text width=25em, text centered, rounded corners, minimum height=3em]
\scalebox{0.7}{
\begin{tikzpicture}[node distance=1.6cm]

\node (n1) at (0,0) [block] {Initialize surface particles at \mbox{$t=0$}: $\mathbf{x}_{i_s}^0, c_{i_s}^{[s]0},\mathbf{n}_{i_s}^0$};
\node (n2) [block, below of=n1, yshift=-0.25cm] {\textbf{while} \mbox{$t<t_\text{end}$}};
\node (n3) [block, below of=n2, xshift=6cm, yshift=-0.25cm]{Evaluate constitutive equation to obtain surface particle velocity field $\mathbf{u}_{i_s}^k=F(c_{i_s}^{[s]k})$};
\node (n4) [block, below of=n3,yshift=-0.75cm] {(Re-)seed bulk particles $\mathbf{x}_{i_b}^k$, initialize their known level-set values $\phi_{i_b}^k$, extend the surface particle velocities to the bulk particles $\mathbf{u}_{i_b}^k$};
\node (n5) [block, below of=n4, yshift=-1.2cm] {Advance $c^{[s]}_{i_s}$ and $\mathbf{x}_{i_s},\mathbf{x}_{i_b}$ to \mbox{$k+1$}:\\
$c_{i_s}^{[s]k+1}=c_{i_s}^{[s]k}+\Delta t\left.\frac{\mathrm{D}c^{[s]}}{\mathrm{D}t}\right|^k$,\\
$\mathbf{x}_{i_b}^{k}=\mathbf{x}_{i_b}^k+\Delta t\mathbf{u}_{i_b}^k$,\\
$\mathbf{x}_{i_s}^{k}=\mathbf{x}_{i_s}^k+\Delta t\mathbf{u}_{i_s}^k$};
\node (n6) [block, below of=n5, yshift=-1.2cm] {Compute geometric information on advected level-set function of bulk particles, evaluate surface normals at surface particle locations $\mathbf{n}_{i_s}^{k+1}$, increment $k\rightarrow k+1$};
%\node (n7) [block, below of=n6, yshift=-0.75cm] {};

% Connectors
\draw [->] (n1) -- (n2);
\draw [->] (n2.south) -| ++(0,-1) |- (n3.west);
\draw [->] (n3) -- (n4);
\draw [->] (n4) -- (n5);
\draw [->] (n5) -- (n6);
\draw [->] (n6.west) -| ++(-6.25,0) |- (n2.west);

\end{tikzpicture}
}
    \caption{Explicit time integration for surface-only PDEs on deformable surfaces.}
    \label{fig:timeIntegrationSurfaceOnly}
\end{figure}
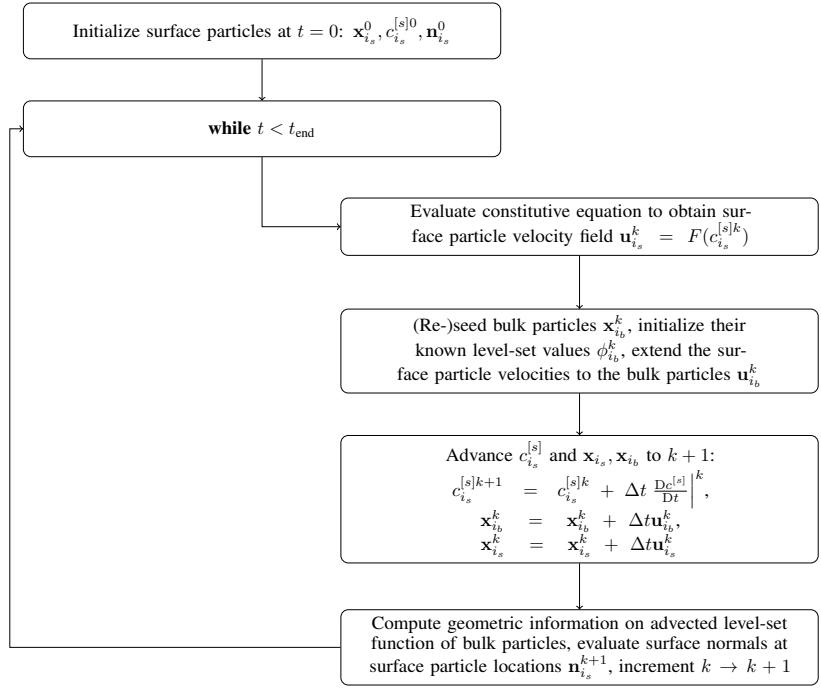

If there is only a surface PDE, and no surrounding bulk fluid, the bulk particles are solely required in a narrow band for the level-set description of the surface. We then evolve the system in time by explicit time integration, as shown in Fig.~\ref{fig:timeIntegrationSurfaceOnly}.

At the beginning of each time step $k$, the surface particle locations $\mathbf{x}_{i_s}^k$, surface concentrations $c_{i_s}^{[s]k}$, and surface normals $\mathbf{n}_{i_s}^k$ at the surface particle locations are known. We use these fields to compute the deformation velocity $\mathbf{u}_{i_s}^k=F(c_{i_s}^{[s]k})$ by evaluating Eq.~\ref{eq:appConstitutiveVel} for each surface particle. We then (re-)seed bulk particles in the narrow band by copying surface particles along their respective surface normal with a fixed spacing $h_s$. Since for each bulk particle the distance to the surface is known by construction, their level-set values $\phi_{i_b}^k$ are initialized analytically to represent a signed-distance function. The velocity of a bulk particle is equal to the velocity of the corresponding surface particle, $\mathbf{u}(\mathbf{x}_{i_b})=\mathbf{u}_{i_b}=\mathbf{u}(\mathbf{cp}(\mathbf{x}_{i_b}))$. 

After initializing the bulk particles, we use the locations, surface normals, and velocities of the surface particles at time $t^k$ to compute the per-particle change in concentration fields $\left.\frac{\mathrm{D}c_{i_s}^{[s]}}{\mathrm{D}t}\right|^k$ by solving Eq.~\eqref{eq:appPDE} using surface DC-PSE. This change is integrated in time using the explicit Euler method,  
\begin{equation}\label{eq:appConcExplEuler}
    c^{[s]k+1}_{i_s}=c^{[s]k}_{i_s} + \Delta t\left.\frac{\mathrm{D}c^{[s]}_{i_s}}{\mathrm{D}t}\right|^k.
\end{equation}
The locations of all particles are then also updated using explicit Euler,
\begin{align}
\mathbf{x}_{i_s}^{k+1}=\mathbf{x}_{i_s}^k+\Delta t~\mathbf{u}_{i_s}^k,\\
\mathbf{x}_{i_b}^{k+1}=\mathbf{x}_{i_b}^k+\Delta t~\mathbf{u}_{i_b}^k.
\end{align}
This advects the level-set function $\phi$, modeling surface movement and deformation. We use the advected level-set function and the PCP method to recompute the surface normals, $\mathbf{n}_{i_s}^{k+1}$, at the new surface particle positions. Finally, using the updated surface normals, the bulk particles are re-seeded orthogonally from the surface particles. Optionally, the SAISS method can be called to regularize the surface particle distribution before re-seeding bulk particles. This completes the time step and provides all information needed for the next time step. The time-step size $\Delta t$ should fulfill the CFL-like condition \cite{schrader2012choosing} for the chemical species with the maximum diffusion constant,
\begin{equation}
    \Delta t\leq2\frac{h_s^2}{\max_s(D^{[s]})}\, .\label{eq:appDiffCFL}
\end{equation}

\subsection{Time integration for coupled surface+bulk PDEs}\label{sec:timeint}

\begin{figure}
    \centering
    \tikzstyle{block} = [rectangle, draw, text width=25em, text centered, rounded corners, minimum height=3em]
\tikzstyle{block2} = [rectangle, draw, text width=30em, text centered, rounded corners, minimum height=3em]
\tikzstyle{block3} = [rectangle, draw, text width=20em, text centered, rounded corners, minimum height=3em]
\scalebox{0.7}{
\begin{tikzpicture}[node distance=1.6cm]

\node (n1) at (0,0) [block] {Initialize bulk and surface particles at \mbox{$t=0$}: $\mathbf{x}_{i_b}^0,\mathbf{u}_{i_b}^0,\phi_{i_b}^0,\rho_{i_b}^0, \mathbf{x}_{i_s}^0, c_{i_s}^{[s]0},\mathbf{u}_{i_s}^0$, compute geometric information};
\node (n2) [block, below of=n1, yshift=-0.3cm] {\textbf{while} \mbox{$t<t_\text{end}$}};
\node (n5) [block3, below of=n2, yshift=-0.0cm, xshift=4.5cm] {Perform predictor step};
\node (n6) [block2, below of=n5, yshift=-0.0cm,xshift=7cm] {Compute bulk densities $\rho_{i_b}^k$, pressures $P_{i_b}^k$, surface tension force $\mathbf{F}_{i_b}^{(s)k}$};
\node (n56) [block2, below of=n6, yshift=-0.4cm] {Obtain surface normals of surface particles $\mathbf{n}_{i_s}^k$ using PCP, regress bulk velocity field to surface particles to obtain $\mathbf{u}_{i_s}^k$};
\node (n7) [block2, below of=n56, yshift=-1.3cm] {Advance $c_{i_s}^{[s]},\mathbf{x}_{i_s}$ and $\mathbf{x}_{i_b},\mathbf{u}_{i_b}$ to intermediate time step $k+1/2$: \\
$c_{i_s}^{[s]k+1/2}=c_{i_s}^{[s]k}+\frac{\Delta t}{2}\left.\frac{\mathrm{D}c^{[s]}}{\mathrm{D}t}\right|^k$,\\
$\mathbf{x}_{i_b}^{k+1/2}=\mathbf{x}_{i_b}^k+\frac{\Delta t}{2}\mathbf{u}_{i_b}^k$,\\
$\mathbf{u}_{i_b}^{k+1/2}=\mathbf{u}_{i_b}^k+\frac{\Delta t}{2}\left.\frac{\mathrm{D}\mathbf{u}}{\mathrm{D}t}\right|^k$,\\
$\mathbf{x}_{i_s}^{k+1/2}=\mathbf{x}_{i_s}^{k}+\frac{\Delta t}{2}\mathbf{u}_{i_s}^k$
};
\node (n8) [block2, below of=n7, yshift=-1.0cm] {Compute geometric information on advected level-set function: $\phi_{i_b}^k\rightarrow\phi_{i_b}^{k+1/2}$};
\node (n9) [block3, below of=n8, yshift=-0.0cm, xshift=-7cm] {Perform corrector step};
\node (n10) [block2, below of=n9, yshift=-0.0cm,xshift=7cm] {Compute bulk densities $\rho_{i_b}^{k+1/2}$, pressures $P_{i_b}^{k+1/2}$, surface tension force $\mathbf{F}_{i_b}^{(s)k+1/2}$};
\node (n1011) [block2, below of=n10, yshift=-0.3cm] {Obtain surface normals of surface particles $\mathbf{n}_{i_s}^{k+1/2}$ using PCP, regress intermediate velocity from bulk particles to surface particles $\mathbf{u}_{i_s}^{k+1/2}$};
\node (n11) [block2, below of=n1011, yshift=-1.4cm] {Advance $c_{i_s}^{[s]},\mathbf{x}_{i_s}$ and $\mathbf{x}_{i_b},\mathbf{u}_{i_b}$ to time step \mbox{$k+1$}:\\
$c_{i_s}^{[s]k+1}=c_{i_s}^{[s]k}+\Delta t\left.\frac{\mathrm{D}c^{[s]}}{\mathrm{D}t}\right|^{k+1/2}$,\\
$\mathbf{x}_{i_b}^{k+1}=\mathbf{x}_{i_b}^{k}+\Delta t\left(\mathbf{u}_{i_b}^k+\frac{\Delta t}{2}\left.\frac{\mathrm{D}\mathbf{u}}{\mathrm{D}t}\right|^{k+1/2}\right)$,\\
$\mathbf{u}_{i_b}^{k+1}=\mathbf{u}_{i_b}^k+\Delta t\left.\frac{\mathrm{D}\mathbf{u}}{\mathrm{D}t}\right|^{k+1/2}$,\\
$\mathbf{x}_{i_s}^{k+1}=\mathbf{x}_{i_s}^{k}+\Delta t\mathbf{u}_{i_s}^{k+1/2}$.};
\node (n12) [block2, below of=n11, yshift=-1.2cm] {Compute geometric information on advected level-set function: $\phi_{i_b}^{k+1/2}\rightarrow\phi_{i_b}^{k+1}$, increment $k\rightarrow k+1$};

% Connectors
\draw [->] (n1) -- (n2);
\draw [->] (n2.south) -| ++(0,-1) |- (n5.west);
\draw [->] (n5.south) -| ++(0,-1) |- (n6.west);
\draw [->] (n6) -- (n56);
\draw [->] (n56) -- (n7);
\draw [->] (n7) -- (n8);
\draw [->] (n8.south) -| ++(0,-1) |- (n9.east);
\draw [->] (n9.south) -| ++(0,-1) |- (n10.west);
\draw [->] (n10) -- (n1011);
\draw [->] (n1011) -- (n11);
\draw [->] (n11) -- (n12);
\draw [->] (n12.west) -| ++(-10.75,0) |- (n2.west);

\end{tikzpicture}
}
    \caption{Predictor-corrector time integration for coupled bulk+surface equations with deformable surfaces.}
    \label{fig:timeIntegration}
\end{figure}
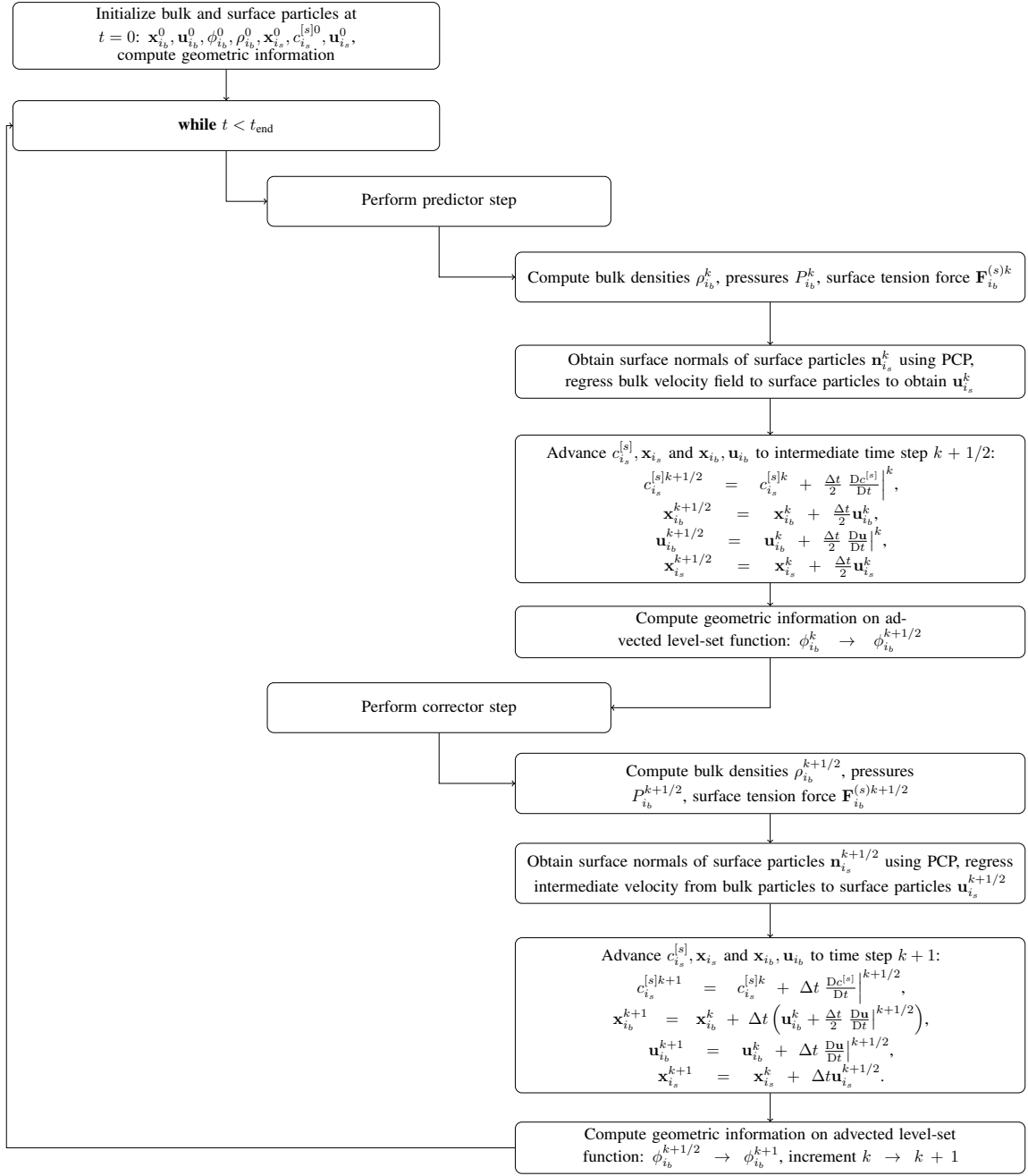

For coupled surface+bulk dynamics, the role of the bulk particles is not purely geometric, but they discretize the bulk fluid according to Eqs.~~\eqref{eq:appmasscontieq}-\eqref{eq:appmomentumeq}. We then evolve both sets of particles over time. Since coupled surface and bulk dynamics can lead to a stiff systems, we use an explicit second-order predictor-corrector scheme \cite{zago2018semi} for enhanced time-integration stability. An overview of the algorithm is shown in Fig.~\ref{fig:timeIntegration}.

At the beginning of each time step $k$, all field values are known and the high-order regression polynomials describing the surface are given. We enter the predictor phase of the time integration by computing the bulk densities $\rho_{i_b}^k$ using Eq.~\eqref{eq:densitysummation} and the bulk pressures $P_{i_b}^k$ using Eq.~\eqref{eq:appEoS}. We then compute the surface tension forces $\mathbf{F}_{i_b}^{(s)k}$ using the geometric information provided by PCP. We also use PCP to compute the surface normals $\mathbf{n}_{i_s}^k$ at the surface particles.

To compute the surface velocity field $\mathbf{u}_{i_s}^k$, we start from the velocity field $\mathbf{u}_{i_b}^{k}$ of the bulk particles within the narrow band around the surface (i.e., for all $i_b$ with $|\phi _{i_b}^k| < w$). We then use local polynomial regression on these particles to obtain the continuous velocity field $\mathbf{u}^{k}(\mathbf{x})$ in the proximity of the surface. This regression problem is identical to the one PCP solves for the level-set function $\phi$ \cite{schulze2024high}, just with a different right-hand side. Therefore, we re-use the orthogonal decomposition from PCP, reducing the computational cost of this step to a single matrix-vector multiplication. From the local regression polynomials, we evaluate the velocity field at the locations of the surface particles to obtain $\mathbf{u}^{k}_{i_s}$.

Using this $\mathbf{u}_{i_s}^k$, the right-hand sides of both the surface and bulk PDEs are evaluated using surface DC-PSE operators on the surface and SPH operators in the bulk. The resulting Lagrangian rates of change are then used to advance the fields and particle locations to an intermediate time step $k+1/2$:
\begin{align}
    &c_{i_s}^{[s]k+1/2}=c_{i_s}^{[s]k}+\frac{\Delta t}{2}\left.\frac{\mathrm{D}c_{i_s}^{[s]}}{\mathrm{D}t}\right|^k,\\
    &\mathbf{x}_{i_b}^{k+1/2}=\mathbf{x}_{i_b}^k+\frac{\Delta t}{2}\mathbf{u}_{i_b}^k,\\
    &\mathbf{u}_{i_b}^{k+1/2}=\mathbf{u}_{i_b}^k+\frac{\Delta t}{2}\left.\frac{\mathrm{D}\mathbf{u}_{i_b}}{\mathrm{D}t}\right|^k,\\
    &\mathbf{x}_{i_s}^{k+1/2}=\mathbf{x}_{i_s}^{k}+\frac{\Delta t}{2}\mathbf{u}_{i_s}^k.\label{eq:appInterVelo}
\end{align}
After this predictor step, PCP is used to re-initialize the level set and recompute the surface normals. 

This is followed by the corrector step, for which we again first compute the bulk densities $\rho_{i_b}^{k+1/2}$ and pressures $P_{i_b}^{k+1/2}$ using Eqs.~\eqref{eq:densitysummation} and \eqref{eq:appEoS}, followed by updating the surface tension $\mathbf{F}_{i_b}^{(s)k+1/2}$. The bulk velocity field is then again regressed to the surface particles by re-using the PCP polynomials. We then evaluate the right-hand sides of the surface and bulk PDEs at the intermediate time step to perform the corrector step to $k+1$,
\begin{align}
    &c_{i_s}^{[s]k+1}=c_{i_s}^{[s]k}+\Delta t\left.\frac{\mathrm{D}c_{i_s}^{[s]}}{\mathrm{D}t}\right|^{k+1/2},\\
    &\mathbf{x}_{i_b}^{k+1}=\mathbf{x}_{i_b}^{k}+\Delta t\left(\mathbf{u}_{i_b}^k+\frac{\Delta t}{2}\left.\frac{\mathrm{D}\mathbf{u}_{i_b}}{\mathrm{D}t}\right|^{k+1/2}\right),\\
    &\mathbf{u}_{i_b}^{k+1}=\mathbf{u}_{i_b}^k+\Delta t\left.\frac{\mathrm{D}\mathbf{u}_{i_b}}{\mathrm{D}t}\right|^{k+1/2},\\
    &\mathbf{x}_{i_s}^{k+1}=\mathbf{x}_{i_s}^{k}+\Delta t\mathbf{u}_{i_s}^{k+1/2}.
\end{align}
This uses a second-order correction for the bulk particle locations \cite{suchde2018point}. Finally, the PCP method is used to re-initialize the level set and recompute the surface normals. The new surface particle distribution can optionally be regularized using the SAISS method as described in Sec.~\ref{sec:saiss}. This completes the time step. 
The time-step size $\Delta t$ should be small enough to fulfill both the CFL-like conditions of the solver of the surface PDE as given in Eq.~\eqref{eq:appDiffCFL} and of the solver for the bulk PDEs. For the present SPH discretization and the predictor-corrector scheme, the latter is
\begin{equation}
    \Delta t \leq\left\{0.25\frac{\epsilon}{c_{ss}+\|\mathbf{u}_{\max}\|_2},~0.25\frac{\rho\epsilon^2}{\eta},~0.25\sqrt{\frac{\rho\epsilon_1^3}{2\pi\tau}}\right\}\, ,
\end{equation}
combining the speed-of-sound condition, viscous condition, and surface-tension condition.

The intermediate predictor step ensures minimal lag between the explicit and the implicit interpretation of the surface representation. Nevertheless, due to numerical errors in the polynomial regression of the surface particle velocities, it can happen that surface particles deviate from the zero level-set over time. This is easily diagnosed by computing the distance of the surface particles to the zero level-set and equally easily corrected by projecting them to their closest point on the surface. Both the distance to the surface and the location of the closest in-surface point are computed anyway when using PCP to re-initialize the level set, so no additional computational cost is incurred. In the results shown in this paper, however, this was never needed and therefore not done.

\section{Results}\label{sec:results}

We implement the above algorithm in C++ using the open-source OpenFPM middleware for scalable scientific computing \cite{incardona2019openfpm}. PCP, SAISS, and surface DC-PSE are all available as modules in the OpenFPM numerics library. We start by presenting results for conservation and convergence benchmarks on simple  problems with known analytical solutions. Then, we present results for more complex models of coupled bulk-and-surface dynamics with emergent interface deformations. We first simulate nonlinear reaction-diffusion patterns on passively relaxing droplet before presenting results for a two-way coupled morphogenetic model.

\subsection{Operator convergence on the unit sphere}

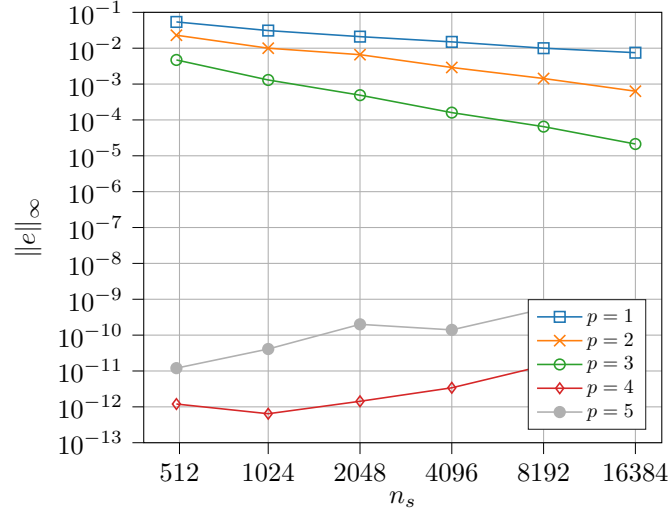
\begin{figure}
    \centering
    % This file was created with tikzplotlib v0.10.1.
\begin{tikzpicture}

\definecolor{crimson2143940}{RGB}{214,39,40}
\definecolor{darkgray176}{RGB}{176,176,176}
\definecolor{darkorange25512714}{RGB}{255,127,14}
\definecolor{forestgreen4416044}{RGB}{44,160,44}
\definecolor{steelblue31119180}{RGB}{31,119,180}

\begin{axis}[
log basis x={10},
log basis y={10},
tick align=outside,
tick pos=left,
x grid style={darkgray176},
xlabel={$n_s$},
xmajorgrids,
xmin=400, xmax=20000,
xmode=log,
xtick style={color=black},
y grid style={darkgray176},
ylabel={$\|e\|_\infty$},
ylabel style={yshift=0.3cm},
ymajorgrids,
ymin=1.0e-13, ymax=0.1,
ymode=log,
ytick style={color=black},
xtick={524,1024,2048,4096,8192,16384},
xticklabels={
\(\displaystyle {512}\),
\(\displaystyle {1024}\),
\(\displaystyle {2048}\),
\(\displaystyle {4096}\),
\(\displaystyle {8192}\),
\(\displaystyle {16384}\),
},
ytick={1e-13,1e-12,1e-11,1e-10,1e-09,1e-08,1e-07,1e-06,1e-05,0.0001,0.001, 0.01, 0.1},
yticklabels={
  \(\displaystyle {10^{-13}}\),
  \(\displaystyle {10^{-12}}\),
  \(\displaystyle {10^{-11}}\),
  \(\displaystyle {10^{-10}}\),
  \(\displaystyle {10^{-9}}\),
  \(\displaystyle {10^{-8}}\),
  \(\displaystyle {10^{-7}}\),
  \(\displaystyle {10^{-6}}\),
  \(\displaystyle {10^{-5}}\),
  \(\displaystyle {10^{-4}}\),
  \(\displaystyle {10^{-3}}\),
  \(\displaystyle {10^{-2}}\),
  \(\displaystyle {10^{-1}}\)
},
legend style={legend cell align=left, align=left, draw=white!15!black, nodes={scale=0.75, transform shape}, mark options={scale=1.1}},
legend pos=south east
]
\addplot [semithick, steelblue31119180, mark=square, mark size=2, mark options={solid}]
table {%
512 5.4e-2
1024    3.1e-2
2048    2.1e-2
4096    1.5e-2
8192    1e-2
16384   7.5e-3
};
\addlegendentry{$p=1$};

\addplot [semithick, darkorange25512714, mark=x, mark size=3, mark options={solid}]
table {%
512 2.3e-2
1024    1e-2
2048    6.6e-3
4096    2.9e-3
8192    1.43e-3
16384   6.33e-4
};
\addlegendentry{$p=2$}

\addplot [semithick, forestgreen4416044, mark=o, mark size=2, mark options={solid}]
table {%
512 4.7e-3
1024    1.3e-3
2048    4.9e-4
4096    1.6e-4
8192    6.5e-5
16384   2.13e-5
};
\addlegendentry{$p=3$}

\addplot [semithick, crimson2143940, mark=diamond, mark size=2, mark options={solid}]
table {%
512 1.2e-12
1024    6.43e-13
2048    1.43e-12
4096    3.4e-12
8192    1.4e-11
16384   3.1e-11
};
\addlegendentry{$p=4$}

\addplot [semithick, darkgray176, mark=*, mark size=2, mark options={solid}]
table {%
512 1.2e-11
1024    4.1e-11
2048    2.0e-10
4096    1.4e-10
8192    5.4e-10
16384   1.6e-10
};

\addlegendentry{$p=5$}

\end{axis}

\end{tikzpicture}
    \caption{Convergence of surface DC-PSE interpolation for the spherical harmonic $Y_{3,2}$ on the unit sphere. The particles from which the function is interpolated are distributed on a Fibonacci sphere grid. The particles set to which the data is interpolated comprises 256 particles distributed on a Fibonacci grid. The cutoff radii for the different operator orders are $r_\text{DCPSE}=\{1.5h_s,1.75h_s,2.25h_s,2.75h_s,3.5h_s\}$, respectively. The empirical convergence orders are $p_\text{emp}=\{1.13,2.06,3.12\}$ for the operator orders $p=\{1,2,3\}$, respectively. For operator orders $p=4$ and $p=5$ the error from the linear system solver dominates, as bounded by the condition number of the system matrix.}
    \label{fig:SDCPSEP2P}
\end{figure}

To verify convergence of the surface DC-PSE interpolation scheme, we discretize the unit sphere with a varying number $n_s$ of surface particles, which are distributed on the sphere according to the Fibonacci sphere sequence \cite{gonzalez2010measurement}, ensuring homogeneous particle neighborhoods. For $n_s$ particles at locations $\mathbf{x}_{i_s}=(x_{i_s},\, y_{i_s},\, z_{i_s})$, we assign values from the spherical harmonic
\begin{equation}
    Y_{3,2}(\mathbf{x}_{i_s}) = \frac{1}{4}\sqrt{\frac{105}{\pi}}(x_{i_s}^2-y_{i_s}^2)z_{i_s},
\end{equation}
and perform particle-to-particle interpolation to a set of 256 Fibonacci particles on the same sphere. We use different surface DC-PSE operator orders and report the maximum absolute error in Fig.~\ref{fig:SDCPSEP2P}. We find close-to-theoretical empirical convergence rates $p_\text{emp}=\{1.13,2.06,3.12\}$ for the operator orders $p=\{1,2,3\}$. For higher-order surface DC-PSE operators, error magnitudes are bounded by the PCP solver tolerance of $10^{-9}$. This is because DC-PSE operators of order $\geq4$ include cubic terms in their basis and the function to be interpolated is $L_1$-cubic. With operators of order 4 or higher, interpolation is therefore exact, while the direct solver of the linear DC-PSE kernel systems has an error proportional to the condition number of the matrix, which is higher for $p=5$ than for $p=4$.

\subsection{Convergence for a passive scalar on a growing sphere}

After testing convergence of the surface DC-PSE operators, we next test the coupling between PCP and surface DC-PSE. For this, we consider a passive scalar field $c$ on the surface of a sphere $\Gamma(t)$ with isotropically growing radius, hence
\begin{align}
\left.\begin{aligned}
\label{eq:growgov}
    &\frac{\mathrm{D}c}{\mathrm{D}t}=-c~\nabla_\Gamma\cdot\mathbf{u}\, ,\\
    &\mathbf{u}=\mathbf{n}
    \end{aligned}\quad\right\}\quad\text{~in~}\Gamma(t)\, .
\end{align}
Since there are no reactions, the total mass on the surface should remain constant while the concentration on the growing surface uniformly decreases.

We initialize the surface to be the unit sphere embedded in $\mathbb{R}^3$. It is represented using bulk particles located on a Cartesian sub-grid with spacing $h_b=1/48$ within a narrow band of width $w=12h_b$ from the surface. The level-set values are initialized to the known analytical distance to the unit sphere. We do not assume the surface normals to be known and instead compute them using PCP  with a cutoff radius $r_c=2.4h_b$ and solver tolerance $\varepsilon_\text{PCP}=10^{-14}$. We then obtain surface particles on the unit sphere using SAISS with $h_s\in\{\frac{1}{8},\frac{1}{16},\frac{1}{32},\frac{1}{64}\}$. All other parameter values are as given in Secs.~\ref{sec:pcp} and \ref{sec:saiss}.

\begin{figure}
    \centering
    % This file was created with tikzplotlib v0.10.1.
\begin{tikzpicture}

\definecolor{crimson2143940}{RGB}{214,39,40}
\definecolor{darkgray176}{RGB}{176,176,176}
\definecolor{darkorange25512714}{RGB}{255,127,14}
\definecolor{forestgreen4416044}{RGB}{44,160,44}
\definecolor{steelblue31119180}{RGB}{31,119,180}

\begin{axis}[
log basis x={10},
log basis y={10},
tick align=outside,
tick pos=left,
x grid style={darkgray176},
xlabel={$h_s$},
xmajorgrids,
xmin=0.0125, xmax=0.15,
xmode=log,
xtick style={color=black},
y grid style={darkgray176},
ylabel={$\|e\|_\infty$},
ylabel style={yshift=0.3cm},
ymajorgrids,
ymin=5.0e-6, ymax=0.003,
ymode=log,
ytick style={color=black},
xtick={0.125,0.0625,0.03125,0.015625},
xticklabels={
\(\displaystyle {1/8}\),
\(\displaystyle {1/16}\),
\(\displaystyle {1/32}\),
\(\displaystyle {1/64}\)
},
ytick={1e-06,1e-05,0.0001,0.001},
yticklabels={
  \(\displaystyle {10^{-6}}\),
  \(\displaystyle {10^{-5}}\),
  \(\displaystyle {10^{-4}}\),
  \(\displaystyle {10^{-3}}\),
},
legend style={legend cell align=left, align=left, draw=white!15!black, nodes={scale=0.75, transform shape}, mark options={scale=1.1}},
legend pos=north west
]
\addplot [semithick, steelblue31119180, mark=square, mark size=2, mark options={solid}]
table {%
0.125     0.000902616
0.0625     0.000382454
0.03125    0.000150495
0.015625    6.62743e-05
};
\addlegendentry{$p=1$};

\addplot [semithick, darkorange25512714, mark=x, mark size=3, mark options={solid}]
table {%
0.125     0.000112584
0.0625     4.23962e-05
0.03125    1.29268e-05
0.015625    5.82917e-06
};
\addlegendentry{$p=2$}

\addplot[semithick, black, dashed]
table {
0.125  5.0e-4
0.015625    7.8125e-6
};

\addplot[semithick, black]
table {
0.125  0.002
0.015625    2.5e-4
};
\end{axis}

\end{tikzpicture}
    \caption{Convergence of a mass-conserved passive scalar on a uniformly growing sphere according to Eq.~\ref{eq:growgov}. We report the maximum absolute error with respect to the analytical solution after 1000 time steps of size $\Delta t=10^{-5}$. The solid black line represents first-order convergence; the dashed black line second-order convergence. The empirical convergence orders for surface DC-PSE operators of orders $p=1$ and $p=2$ are $p_\text{emp}=1.26$ and $p_\text{emp}=1.45$, respectively.}
    \label{fig:massconservation}
\end{figure}
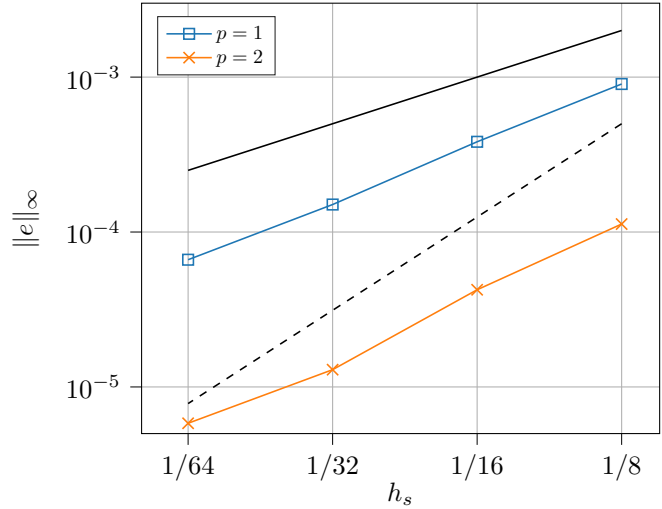

At the surface particles, we initialize $c(\mathbf{x}_{i_s},t_0)=c(\mathbf{x}_{i_s},0)=1.0$. The velocities of the surface particles are constant and equal to the surface normal, see Eq.~\ref{eq:growgov}. We use the time integration scheme from Sec.~\ref{sec:timeintsurfacePDE} for 1000 time steps of size $\Delta t=10^{-5}$. The surface divergence in Eq.~\ref{eq:growgov} is discretized using surface DC-PSE operators of orders $p=1$ and $p=2$ with cutoff radii $r_\text{DCPSE}=1.5h_s$ and $r_\text{DCPSE}=2h_s$, respectively. We report the error with respect to the known analytical solution 
\begin{equation}
    c_\text{theo}(\mathbf{x},t_{1000}) = \left(\frac{1}{\|\mathbf{x}(t_{1000})\|_2}\right)^{\!2}\,c(\mathbf{x},0)\, .
\end{equation}
We report maximum absolute errors after 1000 time steps in Fig.~\ref{fig:massconservation}. We empirical convergence rates of $p_\text{emp}=1.26$ for the first-order operators and $p_\text{emp}=1.45$ for the second-order operators. The maximum errors for $p=2$ are consistently smaller than for $p=1$. The over-convergence for the odd operator order is likely due to error cancellation in odd kernel functions on symmetric domains~\cite{foggia2025numerical}. The under-convergence for $p=2$ is likely due to the first-order Euler time stepping used.

\subsection{Turing patterns on an oscillating fluid droplet}

After verifying the numerical method and its interplay with the geometric computing framework, we consider a more complex case with unknown analytical solution. Specifically, we solve the Gray--Scott reaction-diffusion system \cite{gray1983autocatalytic} on the surface of an viscously relaxing droplet embedded in a bulk fluid. The two surface species $c^{[1]},c^{[2]}$ diffuse independently, but with different diffusion constants, and undergo a third-order nonlinear reaction $c^{[1]}+2c^{[2]}\rightarrow 3c^{[2]}$, while $c^{[1]}$ is  produced with feed rate $F$ and $c^{[2]}$ is removed with kill rate $K$, hence:
\begin{align}\label{eq:appGrayscott}
&\left.\begin{aligned}
&\frac{\mathrm{D}c^{[1]}}{\mathrm{D}t}=D^{[1]}\Delta_\Gamma c^{[1]}-c^{[1]}(c^{[2]})^2+F(1-c^{[1]})-c^{[1]}\nabla_\Gamma\cdot\mathbf{u}\\
&\frac{\mathrm{D}c^{[2]}}{\mathrm{D}t}=D^{[2]}\Delta_\Gamma c^{[2]}+c^{[1]}(c^{[2]})^2-(F+K)c^{[2]}-c^{[2]}\nabla_\Gamma\cdot\mathbf{u}
\end{aligned}\quad\right\}\quad\text{~in~}\Gamma(t)\, .
\end{align}
Depending on the choice of coefficients, this reaction-diffusion system is able to generate and maintain a wide range of non-equilibirum concentration patterns akin to Turing patterns \cite{turing1952chemical}. 
Here, we consider a subset of the parameters from Ref.~\cite{pearson1993complex}, which provides us with a reference solution for the corresponding problem in a flat 2D space. Specifically, we set the diffusion constants to $D^{[1]}=0.0002,\, D^{[2]}=0.0001$ and consider the $\lambda$ ($F=0.036,\, K=0.066$), $\gamma$ ($F=0.024,\, K=0.056$), and $\alpha$ ($F=0.016,\, K=0.05$) patterns. We rescale the right-hand side of Eq.~\eqref{eq:appGrayscott} with $1/\Delta t$ such that the physical time scale of pattern formation is comparable with the time scale of the oscillation of the fluid droplet, rendering them coupled. Both processes then reach a steady state within the total simulation time $t_\text{end}=2$.

The deformation velocity $\mathbf{u}$ in Eq.~\eqref{eq:appGrayscott} is computed as the solution of the bulk Eqs.~\eqref{eq:appmasscontieq} and \eqref{eq:appmomentumeq}. 
The volumetric surface tension force $\mathbf{F}^{(s)}$ in Eq.~\eqref{eq:csfmodel} acts in the proximity of the surface of the droplet, which initially is an ellipsoid with major axes $A=0.75,\,B=C=0.5$ and surface-tension coefficient  $\tau=10$. The incompressible hydrodynamic phases surrounding the surface have a reference density of $\rho_0=1$, an artificial speed of sound of $c_{ss}=70$, a polytropic index of $\gamma=7$, and a viscosity of $\eta=0.05$. The bulk particles initially have zero velocity $\mathbf{u}_{i_b}=(0,0,0)^\top$ and are only accelerated by the tension of the relaxing droplet surface.

To prevent the system from relaxing to the trivial solution $c^{[1]}=1,c^{[2]}=0$, we initialize the surface species in two polar patches $|x|>0.625 \land |y|<0.125 \land |z|<0.125$ as $c^{[1]}=0.5, \,c^{[2]}=0.25$, and we add initial noise in the rest of the surface as $c^{[1]}=0.9+\mu$, $c^{[2]}=\mu$. Here, $\mu\sim\mathcal{U}[0,\, 0.1]$ is a pseudo-random variable {\it i.i.d.} from the uniform distribution over the closed interval $[0,\,0.1]$. We place bulk particles in the entire embedding domain initially on a Cartesian grid with inter-particle spacing $h_b=1/32$. For the level-set description of the surface, we use a narrow band of width $w=10h_b$, a PCP cutoff radius of $r_c=3h_b$, and a solver tolerance of $\varepsilon_\text{PCP}=10^{-10}$. We choose a smoothing length of $\epsilon=\epsilon_{1}=2h_b$ for the SPH operators in the bulk and order $p=2$, cutoff radius $r_\text{DCPSE}=2.5h_s$, and extension spacing $h_s=h_b$ for the DC-PSE operators on the surface. Time stepping is done as described in Sec.~\ref{sec:timeint} with step size $\Delta t=2\times10^{-4}$, which fulfills all CFL-like conditions of the SPH solver and the CFL-like condition in Eq.~\eqref{eq:appDiffCFL}. We evolve the system without intermediate surface particle regularization until $t_\text{end}=2$.

The initial surface particle set is obtained in one of two ways: (1) By using the sample particles of the PCP method. Due to the curvature gradients on the ellipsoid, this distribution is inhomogeneous with spatially varying nearest-neighbor distances. (2) By using SAISS with (1) as initial condition to obtain surface particle cloud with spatially homogeneous resolution of $h_s=h_b=1/32$. This results in 4470 surface particles on the initial ellipsoid droplet.

\begin{figure}
    \centering
    \includegraphics[width=\textwidth]{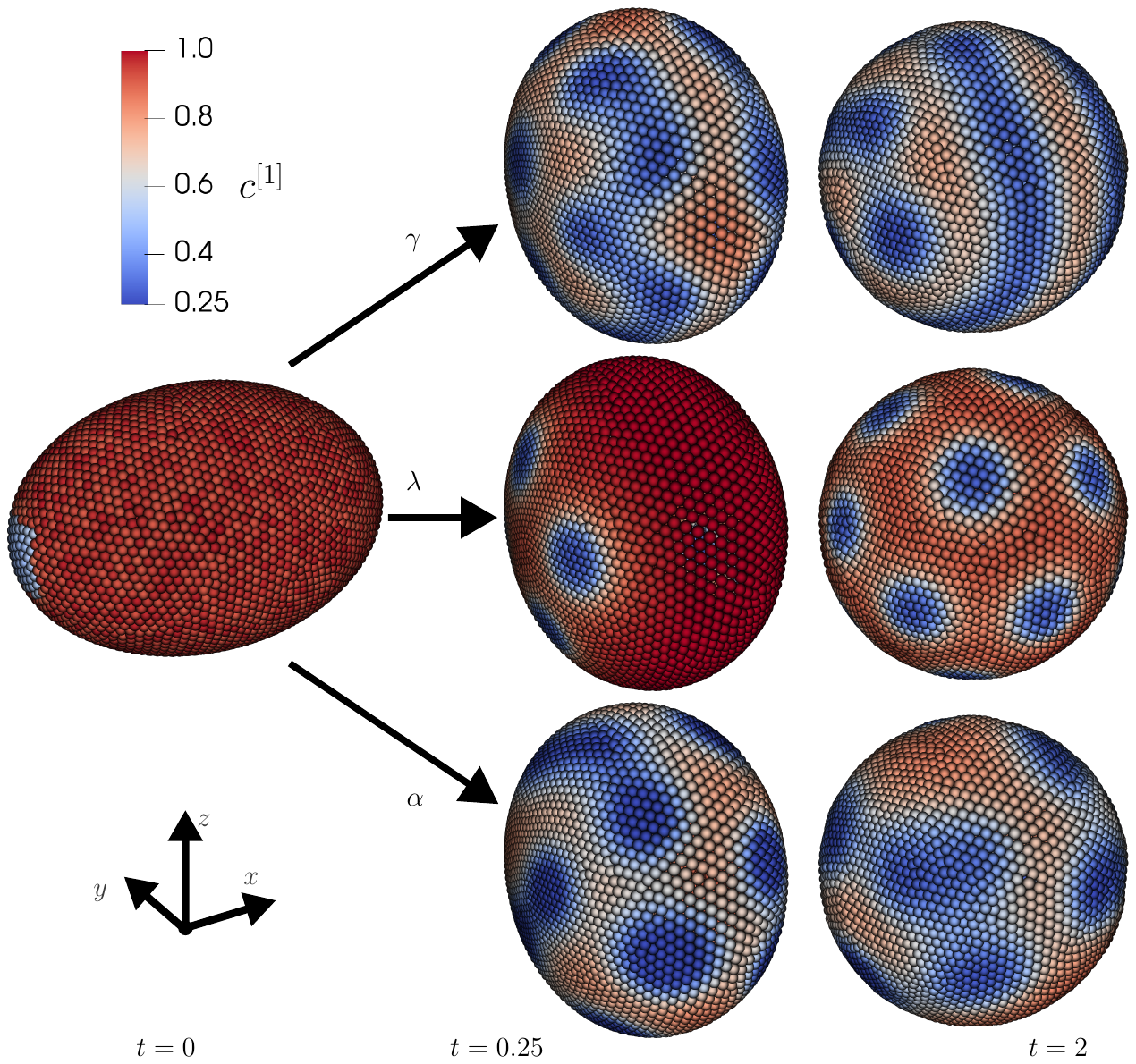}
    \caption{Surface particles of the 3D oscillating droplet for three different parameter configurations ($\gamma,\lambda,\alpha$) of the Gray--Scott reaction-diffusion system within the dynamic surface. The surface particles are color-coded by the concentration $c^{[1]}$ (color bar). Their positions and concentrations are shown for three time points: Initially at $t=0$, during the maximum vertical extension of the droplet at $t=0.25$, and when the droplet shape has equilibrated at $t=2$.}
   \label{fig:3DdropletSurfaceRD}
\end{figure}

We visualize the locations and concentrations $c^{[1]}$ of the surface particles from initialization (2) in Fig.~\ref{fig:3DdropletSurfaceRD} at three time points of the simulation. The Gray--Scott patterns are qualitatively similar to the flat 2D case \cite{pearson1993complex}: For the $\gamma$ pattern, we find a steady state mainly consisting of lines. The $\lambda$ pattern forms slower and yields steady dots evenly distributed over the surface of the droplet. The $\alpha$ pattern does not have a steady state and instead generates a dynamic distribution of indefinitely moving, fusing, and splitting dots on the surface.

\begin{figure}
    \centering
    \includegraphics[width=0.6\textwidth]{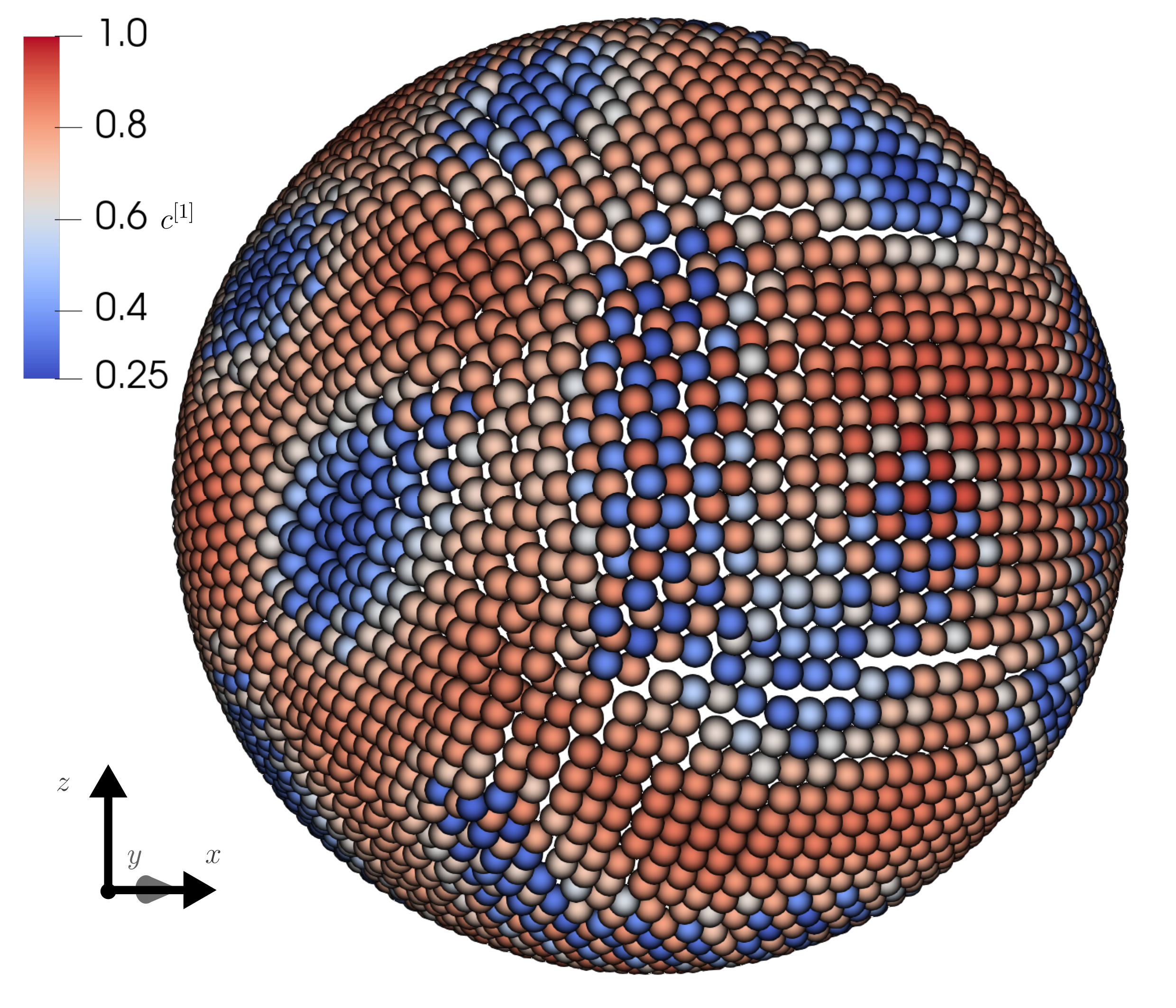}
    \caption{Surface particles of the 3D oscillating droplet at steady state for a higher viscosity $\eta=0.2$ at $t=1$. In this simulation, the initial surface particle distribution was not approximately equidistant. The surface particles are color-coded according to their concentration $c^{[1]}$ (color bar). The non-physical checkerboard pattern heralds the onset of a fatal numerical instability when not using sufficiently regular surface particle distributions.}
    \label{fig:3DdropletSurfaceRDCheckerboard}
\end{figure}

The regularity of the surface particle distribution obtained using the SAISS method is crucial for stable and consistent simulations. This is immediately seen when starting the same simulation from the initial surface particle distribution obtained using approach (1), without using SAISS. This quickly yields non-physical results. At $t\approx0.12$, the surface concentration $c^{[1]}$ becomes non-smooth with large positive ($c^{[1]}=8.6$) and negative ($c^{[1]}=-9.1$) values at adjacent particles. At $t\approx0.125$, the field values leave the space of machine numbers, indicating a fatal numerical instability. A higher viscosity of the surrounding bulk fluid delays the onset of the instability. For $\eta=0.2$, the droplet shape is able to equilibrate to an unphysical, yet stable steady state (Fig.~\ref{fig:3DdropletSurfaceRDCheckerboard}). The checkerboard-like pattern of the concentration field is then clearly visible, independent of surface deformation. In all simulations using the homogeneous initial particle distribution obtained using SAISS, the resulting final fields were smooth and stable, emphasizing the importance of surface-discretization regularity when solving PDEs on dynamic surfaces.

\subsection{Fully coupled morphogenetic model}

While the previous example combined nonlinear in-surface dynamics with surrounding bulk fluids to govern the surface deformation, the coupling was only one way. We next test full two-way coupling between surface deformation and in-surface dynamics by considering a simple model of biological morphogenesis. The reaction-diffusion dynamics within the surface is again governed by the Gray--Scott Eq.~\eqref{eq:appGrayscott}. The surface deformation velocity $\mathbf{u}$ is computed from the concentration field $c^{[2]}$ as
\begin{equation}
    \mathbf{u}=uc^{[2]}\mathbf{n}\, ,
\end{equation}
with the coefficient $u=1.5$ controlling growth speed. For the Gray--Scott system, we set $D^{[1]}=0.0002,\, D^{[2]}=0.0001$ and solve the $\gamma$ and $\alpha$ patterns for a right-hand-side scaling factor of $2.5/\Delta t$ with $\Delta t=10^{-3}$.

Initially, the surface is a sphere of radius $R_0=0.5$. The surface concentration fields are initialized as for the oscillating droplet but with initial patches  at $|x|>0.25 \land |y|<0.25 \land |z|<0.25$. For the PCP level set, we use a narrow band of width $w=10h_b$ with bulk inter-particle spacing $h_b=1/40$. The PCP cutoff radius is $r_c=2.6h_b$ and the solver tolerance $\varepsilon_\text{PCP}=10^{-7}$. We use SAISS, starting from the PCP sample particles, to obtain a regular surface discretization with $h_s=h_b=1/40$. This results in 5048 surface particles initially. With the same parameters, SAISS is used with a frequency of $\mathfrak{f}_\text{SAISS}=20/t$ (i.e., 20 times per unit of dimensionless simulation time) to regularize the surface particles and adjust their total number to the evolving surface area. For comparison, we also run a simulation without periodic re-regularization of the surface particle distribution, i.e., with $\mathfrak{f}_\text{SAISS}=0$. For surface DC-PSE, we choose a cutoff radius of $r_\text{DCPSE}=2.5h_s$ and approximation order $p=2$.

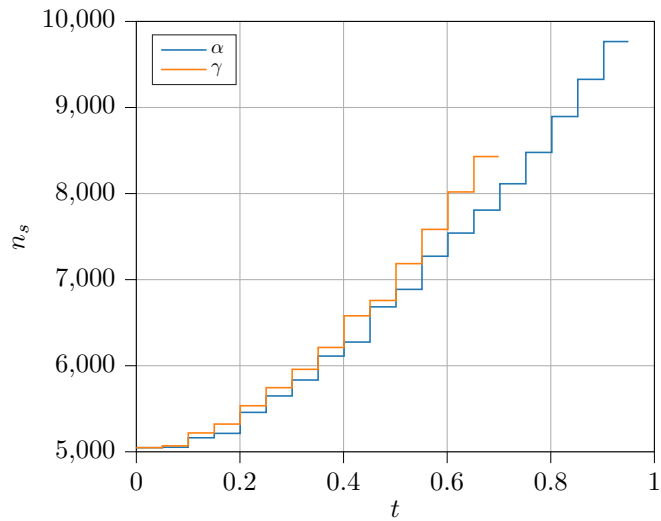
\begin{figure}
    \centering
    \begin{tikzpicture}

\definecolor{darkgray176}{RGB}{176,176,176}
\definecolor{darkorange25512714}{RGB}{255,127,14}
\definecolor{steelblue31119180}{RGB}{31,119,180}

\begin{axis}[
tick align=outside,
tick pos=left,
x grid style={darkgray176},
xlabel={$t$},
xmajorgrids,
xmin=0, xmax=1,
xtick style={color=black},
xtick={0,0.2,0.4,0.6,0.8,1.0},
y grid style={darkgray176},
ylabel={$n_s$},
ylabel style={yshift=0.3cm},
ymajorgrids,
ymin=5000, ymax=10000,
ytick style={color=black},
ytick={5000,6000,7000,8000,9000,10000},
scaled y ticks=false,
y tick label style={/pgf/number format/fixed},
legend style={legend cell align=left, align=left, draw=white!15!black, nodes={scale=0.75, transform shape}, mark options={scale=1.1}},
legend pos=north west
]

% --- alpha ---
\addplot [const plot, semithick, steelblue31119180, mark=none, mark size=1.5, mark options={solid}]
table {%
0.00000 5048
0.05013 5053
0.10025 5163
0.15038 5214
0.20050 5458
0.25063 5649
0.30075 5834
0.35088 6112
0.40100 6275
0.45113 6684
0.50125 6887
0.55138 7272
0.60150 7541
0.65163 7808
0.70175 8114
0.75188 8478
0.80201 8895
0.85213 9327
0.90226 9766
0.94987 9766
};
\addlegendentry{$\alpha$}

% --- gamma ---
\addplot [const plot, semithick, darkorange25512714, mark=none, mark size=2.5, mark options={solid}]
table {%
0.00000 5048
0.05013 5069
0.10025 5219
0.15038 5322
0.20050 5535
0.25063 5745
0.30075 5958
0.35088 6212
0.40100 6580
0.45113 6758
0.50125 7186
0.55138 7584
0.60150 8019
0.65163 8430
0.69925 8430
};
\addlegendentry{$\gamma$}

\end{axis}

\end{tikzpicture}
    \caption{Dynamic adaptation of the number of surface particles $n_s$ over simulated time $t$ for two different Gray--Scott reaction-diffusion patterns ($\alpha$, $\gamma$) in the 3D morphogenesis model. The width of each step equals $1/\mathfrak{f}=0.05$.}
    \label{fig:particlesOverTime}
\end{figure}

\begin{figure}
    \centering
    \includegraphics[width=\linewidth]{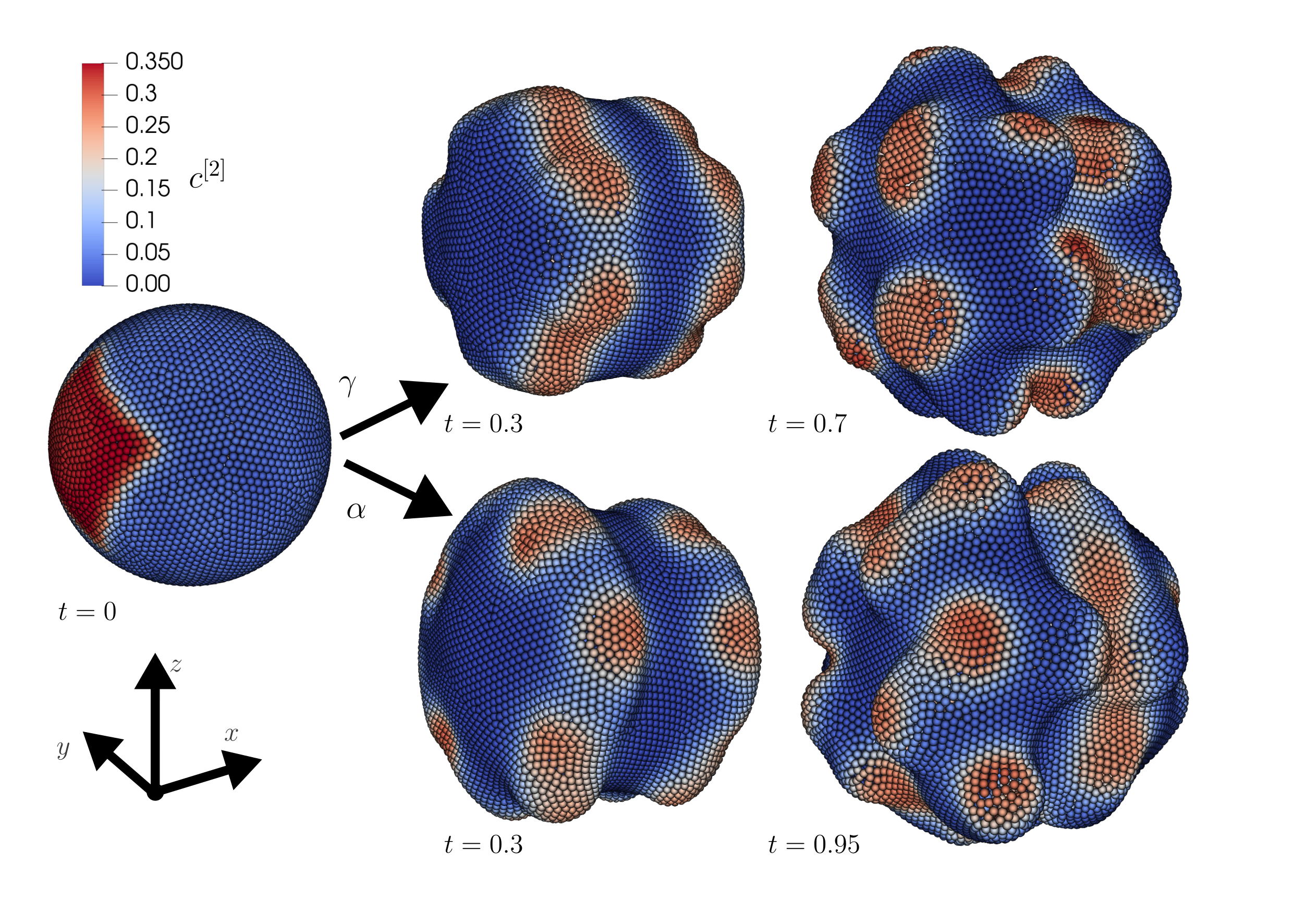}
    \caption{Surface particles for the 3D morphogenesis model with two different Gray--Scott reaction-diffusion patterns ($\gamma,\alpha)$ evolving on the deforming surface. The particles are color-coded according to their surface concentration $c^{[2]}$ (color bar), which induces normal growth of the surface. We show different time points for the two patterns where the resulting shapes are visually similar.}
    \label{fig:morphogenesis}
\end{figure}

The results with periodic particle re-regularization ($\mathfrak{f}_\text{SAISS}=20$) are shown in Fig.~\ref{fig:morphogenesis}. The SAISS method automatically adjusts the total number of particles depending on the dynamically evolving surface area. This is shown in Fig.~\ref{fig:particlesOverTime}. Over the course of the simulation, the number of surface particles increases to 9766 for the $\alpha$ pattern and 8430 for the $\gamma$ pattern, reflecting the different shape evolution of the two morphogenetic patterns. 

For both the $\alpha$ and $\gamma$ patterns, the shape initially remains approximately axially symmetric and then evolves in a way that is reminiscent of tumor or coral growth. The final shape for the $\gamma$ pattern has almost equidistant, roughly spherical lobes, which originated from two lines around the sphere. The $\alpha$ pattern yields a more irregular final shape with asymmetric lobes. Both patterns are qualitatively different from the same parameter sets on the oscillating droplet (Fig.~\ref{fig:3DdropletSurfaceRD}), highlighting the effect of two-way coupling \cite{frank2019pinning}.

We stop the simulations once the largest curvature of the evolving surface has reached the maximum admissible value for the given narrow-band width. Beyond that point, narrow bands from opposing surfaces would start intersecting, leading to numerical instabilities. Surface DC-PSE requires the surface to possess a non-intersecting tubular neighborhood of width $w$ \cite{singh2023meshfree}. This defines the simulation end times $t_\text{end}=0.7$ for the $\gamma$ pattern and $t_\text{end}=0.95$ for the $\alpha$ pattern.

\begin{figure}
    \centering
    \includegraphics[width=\linewidth]{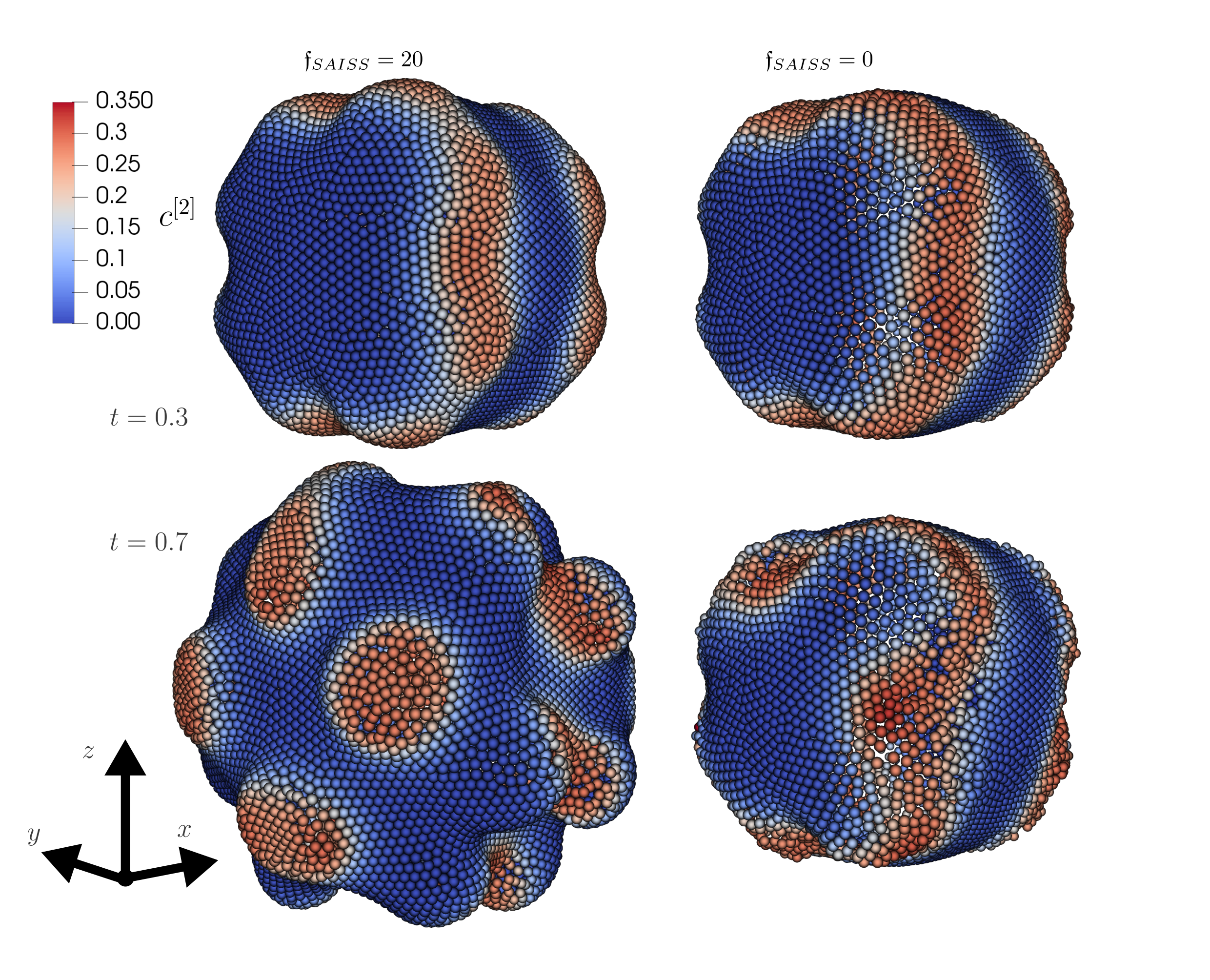}
    \caption{Surface particles for the 3D morphogenesis model with the $\gamma$ pattern with (left column, $\mathfrak{f}_\text{SAISS}=20$) and without (right column, $\mathfrak{f}_\text{SAISS}=0$) periodic re-regularization of the particle positions at two time points ($t=0.3,\,t=0.7$). The particles are color-coded according to their respective surface concentration $c^{[2]}$ (color bar).}
    \label{fig:morphogenesisNoReg}
\end{figure}

To assess the importance of surface particle regularity, we compare the simulation for the $\gamma$ pattern with and without periodic surface particle re-regularization using SAISS. The results in Fig.~\ref{fig:morphogenesisNoReg} show that the systems evolve comparably at early times but qualitatively differ for later times when the Lagrangian distortion of the surface particles becomes significant. In particular, for $\mathfrak{f}_\text{SAISS}=0$, growth stagnates after $\approx$30\% of the simulation time. The surface concentration field remains smooth and reaches a steady state. The surface particle locations, however, start oscillating along the normal to the surface, indicating numerical instability in the PCP solver. Without periodic surface particle re-regularization, the overall growth of the shape is under-estimated by over 40\% in the end. This shows that in order to simulate large deformations and surface area changes, periodic re-regularization of the particle locations is indispensable.

The irregular surface particle distribution from deformations in the absence of periodic re-regularization also negatively affects runtime. Even though the simulation with $\mathfrak{f}_\text{SAISS}=20$ has 16\% more particles at $t=0.3$ than the simulation with $\mathfrak{f}_\text{SAISS}=0$, each call to the PCP method requires only 26\% of the computational time it requires with $\mathfrak{f}_\text{SAISS}=0$. This is because PCP fails to find good seed particles for irregular surface distributions, requiring more solver iterations. The four-fold PCP speedup on regular particle distributions amortizes the additional time required by SAISS, resulting in an overall shorter wall-clock time of the simulation with more accurate results.

\section{Conclusions and outlook}

We have presented a meshfree numerical solver for coupled bulk-surface PDEs with dynamically deforming surfaces where the surface deformation emerges from the  nonlinear dynamics. The presented method leverages a semi-implicit surface representation to compute geometric quantities---such as normals and curvatures---to high order of accuracy in the embedding space, while solving in-surface dynamics on intrinsic surface particles. We presented an explicit predictor-corrector scheme for evolving the coupled bulk-surface dynamics in time and highlighted the importance of periodically regularizing the surface particle distribution, especially when simulating large deformations or surface-area changes.

We verified the convergence of the presented solver in two simple cases with known analytical solutions. First, we verified convergence of surface DC-PSE interpolation of a spherical harmonic for convergence orders up to 5 on the unit sphere. Second, we showed convergence with the theoretically expected order for the conservation of mass on the surface of an isotropically growing sphere.

We then considered a one-way coupled bulk-surface problem, in which a Gray--Scott reaction-diffusion system on the surface of a 3D droplet was embedded in two bulk hydrodynamic phases in the embedding space. The droplet surface oscillated as driven by the viscous relaxation of the surrounding fluids. On this dynamically deforming surface, different Gray--Scott patterns evolved to steady states comparable to the 2D flat case. We found that the solution of the surface PDE was sensitive to the quality of the surface particle distribution. Simulations with initially irregular particle distributions became unstable or displayed non-physical checkerboard patterns.

Finally, we considered the two-way (between surface concentration fields and surface deformation) coupled case of a simple morphogenetic model, in which two chemical species diffuse and react within an initially spherical surface and induce local growth depending on the surface concentrations. This led to rich, dynamically evolving organic shapes. Due to the large deformations in this case, periodic re-regularization of the surface particle locations was mandatory and also led to faster simulation runtimes. 

However, the morphogenesis simulation also exposed the main limitation of the present method: Surface DC-PSE requires a non-intersecting tubular neighborhood of at least the radius of the narrow band used for surface representation. This became violated for later times of the morphogenetic dynamics and would have led to  
non-physical values in the surface concentration field if the simulations were continued beyond that point. 
Relaxing this limitation in the future would require a curvature-adaptive spacing of the surface particles. While curvature-adaptive distributions can readily be provided by SAISS, the DC-PSE operators would need to be extended with multi-resolution neighbor lists, and the PCP method would require adaptive support radii. Adaptive-resolution fast neighbor lists \cite{Awile:2012} are not trivial to implement in a scalable and efficient way for dynamic surfaces.

A further limitation of the proposed solver for coupled bulk-surface PDEs is the lack of feedback from the surface dynamics to the embedding hydrodynamic phases. This could be achieved, for example, by making the surface tension depend on a local surfactant concentration. While the time-integration scheme presented here does extend to such cases, the bulk particles that exert the surface tension forces would need to be associated with a surfactant concentration value. This would require interpolating the surfactant concentration field to the closest points of bulk particles.

Notwithstanding these limitations, the presented solver offers the possibility of meshfree, high-order Lagrangian solutions of nonlinear surface dynamics on dynamically deforming surfaces coupled with bulk fluids. This provides exciting opportunities for studying the continuum mechanics of biological morphogenesis, multi-phase processes in additive manufacturing, and deformable thin-shell objects in computer graphics.

\section*{CRediT authorship contribution statement}
\textbf{LS} Conceptualization, methodology, software, validation, formal analysis, investigation, data curation, writing - original draft, writing - review and editing, visualization. \textbf{AF} Conceptualization, methodology, software, validation, formal analysis, investigation, writing - original draft, writing - review and editing. \textbf{IFS} Conceptualization, methodology, formal analysis, investigation, resources, writing - original draft, writing - review and editing, supervision, project administration, funding acquisition.

\section*{Data availability}

All examples shown in this manuscript are contained in the ``example'' folder of the OpenFPM repository at: \url{https://git.mpi-cbg.de/mosaic/software/parallel-computing/openfpm/openfpm/-/tree/adaptiveSurfaceParticles/example/Numerics/Coupled_BulkSurface}.

\section*{Acknowledgments}
This work has been funded by the German Federal Ministry of Research, Technology and Space (BMFTR, Bundesministerium f\"{u}r Forschung, Technologie und Raumfahrt) under grant 6G-life (ID 16KISK001K) (\textbf{LS}, \textbf{AF}), and the German Research Foundation (DFG, Deutsche Forschungsgemeinschaft) under grant FOR-3013 (``Vector- and tensor-valued surface PDEs'', project number 417223351) (\textbf{AF}).

\end{document}